\documentclass[sigconf, nonacm]{acmart}

\newcommand\vldbdoi{XX.XX/XXX.XX}

\newcommand\vldbavailabilityurl{URL_TO_YOUR_ARTIFACTS}

\usepackage{tikz}
\usepackage{amsmath}
\usepackage{subcaption}
\usepackage{multirow}
\usepackage{pifont}
\usepackage{listings}
\usepackage[linesnumbered, noend, ruled, vlined]{algorithm2e}
\usepackage{xcolor}
\usepackage{subcaption}
\usepackage{multirow}
\usepackage{array}
\usepackage{graphicx}
\usepackage{xcolor}
\usepackage{color}

\definecolor{ZpfGreen}{RGB}{0,100,0}
\definecolor{ZpfRed}{RGB}{255,0,102}

\usepackage{filecontents}

\usepackage{color}
\definecolor{ZpfGreen}{RGB}{0,100,0}
\definecolor{ZpfRed}{RGB}{255,0,102}

\newcommand{\stitlestart}[1]{\noindent{\bf #1.\/}}

\newcommand{\stitle}[1]{\vspace*{0.3em}\noindent{\bf #1.\/}}

\newcommand{\name}{{\textsf{Gorgeous}}}
\newcommand{\evname}{{\textit{Gorgeous}}}

\newcommand{\squishlist}{
	\begin{list}{$\bullet$}
		{ \setlength{\itemsep}{1pt}
			\setlength{\parsep}{1pt}
			\setlength{\topsep}{2.5pt}
			\setlength{\partopsep}{0.5pt}
			\setlength{\leftmargin}{1em}
			\setlength{\labelwidth}{1em}
			\setlength{\labelsep}{0.6em}
		}
	}
	\newcommand{\squishend}{
	\end{list}
}

\newcommand{\modify}[1]{\textcolor{black}{#1}}

\begin{document}

\author{Peiqi Yin$^{1}$, Xiao Yan$^{2}$, Qihui Zhou$^{1}$, Hui Li$^{1}$, Xiaolu Li$^{3}$, Lin Zhang$^{4}$, Meiling Wang$^{4}$, Xin Yao$^{4}$,}
\author{James Cheng$^{1}$}
\affiliation{%
  \institution{$^1$\{pqyin22, qhzhou, hli, jcheng\}@cse.cuhk.edu.hk, $^2$yanxiaosunny@gmail.com, $^3$lixl666@hust.edu.cn, }
  \institution{$^4$\{zhang.lin4, wangmeiling17, yao.xin1\}@huawei.com}
  \institution{$^1$The Chinese University of Hong Kong, $^2$Wuhan University, $^3$Huazhong University of Science and Technology,}
  \institution{$^4$Huawei Theory Lab}
}










\title{Gorgeous: Revisiting the Data Layout for Disk-Resident High-Dimensional Vector Search}

\begin{abstract}

Similarity-based vector search underpins many important applications, but a key challenge is processing massive vector datasets (e.g., in TBs). To reduce costs, some systems utilize SSDs as the primary data storage. They employ a \textit{proximity graph}, which connects similar vectors to form a graph and is the state-of-the-art index for vector search. However, these systems are hindered by sub-optimal data layouts that fail to effectively utilize valuable memory space to reduce disk access and suffer from poor locality for accessing disk-resident data. Through extensive profiling and analysis, we found that \textit{the structure of the proximity graph index} is accessed more frequently than the \textit{vectors} themselves, yet existing systems do not distinguish between the two. To address this problem, we design the \textit{Gorgeous} system with the principle of prioritizing graph structure over vectors. Specifically, Gorgeous features a memory cache that keeps the adjacency lists of graph nodes to improve cache hits and a disk block format that explicitly stores neighbors' adjacency lists along with a vector to enhance data locality. Experimental results show that Gorgeous consistently outperforms two state-of-the-art disk-based systems for vector search, boosting average query throughput by over 60\% and reducing query latency by over 35\%.

\end{abstract}

\maketitle


\ifdefempty{\vldbavailabilityurl}{}{
\vspace{.3cm}
\begingroup\small\noindent\raggedright\textbf{Artifact Availability:}\\
The source code, data, and/or other artifacts have been made available at \url{https://github.com/yinpeiqi/Gorgeous}.
\endgroup
}

\section{Introduction}

To manage data with complex semantics, e.g., texts, images, and videos, a common practice is to map such data to high-dimensional vector embeddings using machine learning models~\cite{bge, img2emb, video2emb}. A fundamental operation on these vectors is nearest neighbor search (NNS), which retrieves the top-$k$ similar vectors to a given query vector (e.g., as measured by Euclidean distance) from a vector dataset. Since exact NNS requires a linear scan in high-dimensional space~\cite{exactknn}, approximate NNS (ANNS) is commonly employed to return most, instead of all, of the top-$k$ similar vectors, trading some accuracy for efficiency. ANNS is crucial for many applications, including recommendation~\cite{ali_search, fb_retrival}, content search (e.g., for images and videos)~\cite{ann4text_retrieval, taobao_retrival}, retrieval-augmented generation (RAG) for large language models (LLMs)~\cite{rag1, rag2, rag3}, and bio-informatics~\cite{bio1, bio2}. 
These applications require ANNS to achieve a high recall for the search results\footnote{Recall measures the quality of ANNS results. If $k'$ of the returned vectors belong to the ground-truth top-$k$ similar vectors for a query, the recall is $k'/k$.} while ensuring low latency and high throughput for query processing. 

Real-world vector datasets can be massive, containing billions of vectors and occupying TBs of space~\cite{laion5b,fusionanns}. Storing all vectors in memory requires a huge DRAM and is expensive. To reduce costs, several systems explore disk-based ANNS~\cite{diskann,starling,spann}, utilizing SSDs as the primary data storage and using a small memory (e.g., 10\%-20\% of the dataset size) as the cache. DiskANN~\cite{diskann} and Starling~\cite{starling} are state-of-the-art systems for disk-based ANNS. They employ a proximity graph index~\cite{hnsw, bang}, where each vector acts as a graph node connected to similar vectors. ANNS is conducted by traversing the graph towards vectors that are more similar to the input query. When visiting a node\footnote{We use node and vector interchangeably in this paper.}, the search process retrieves its neighbors and computes their distances to the query. DiskANN and Starling use memory to store a compressed version of all vectors~\cite{pq, opq, rabitq}, which allows to compute approximate distances to the query, while SSD holds the original vectors and their adjacency lists (i.e., the proximity graph index). When a node is visited, these systems fetch its original vector (i.e., to compute exact distance and rank the node in the search results) and adjacency list (i.e., to specify its neighbors for approximate distance computation) from disk.

\begin{figure}[!t]
	\centering
	\includegraphics[width=1\linewidth]{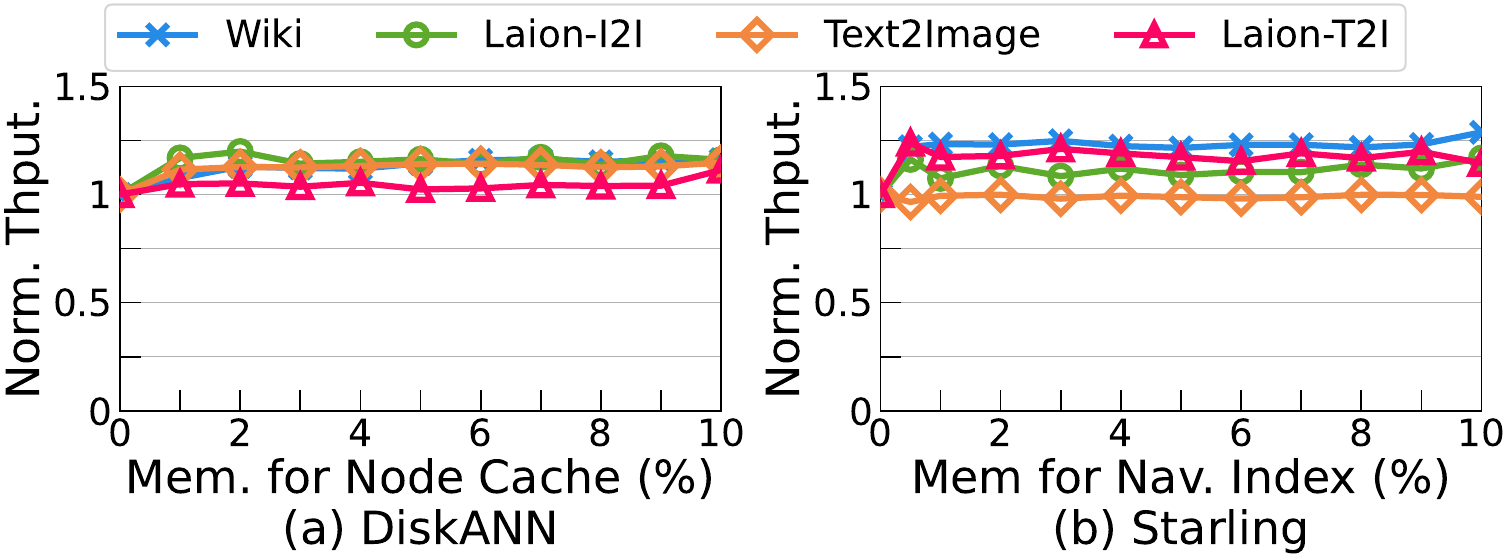}
	\caption{Query throughput of DiskANN and Starling when using memory cache, normalized by the case without memory cache. Cache size is measured by the ratio of dataset size.}
	\label{fig:intro_mem_ratio}
\end{figure}

\begin{figure}[!t]
	\centering
	\includegraphics[width=1\linewidth]{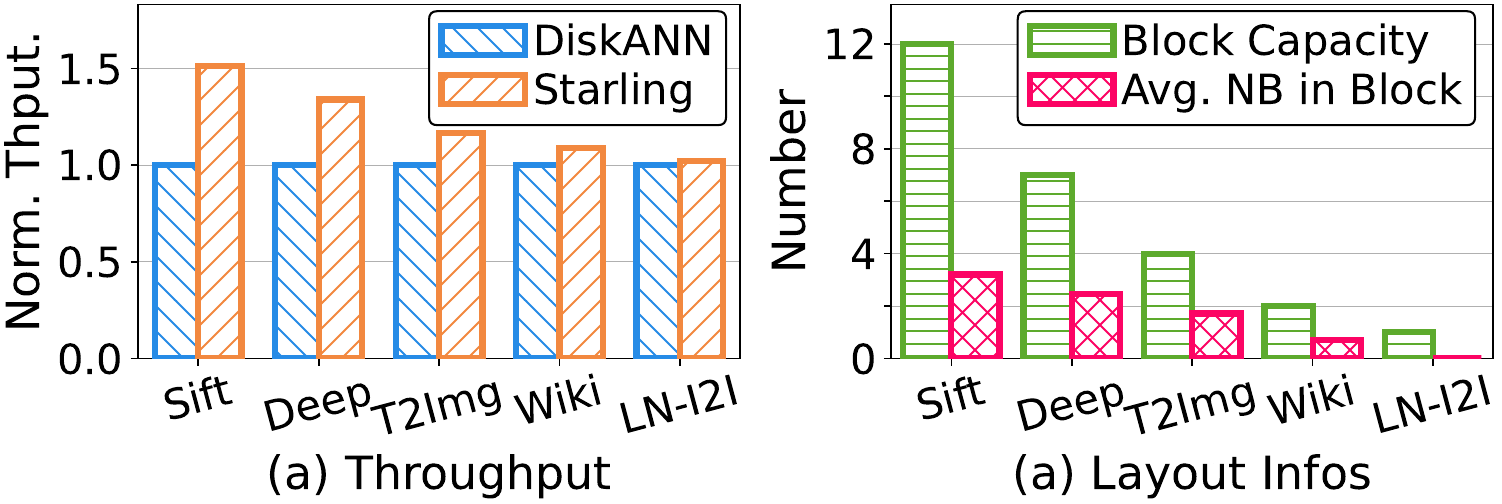}
	\caption{(a) The performance gain of Starling's disk data layout over DiskANN, query throughput is normalized by DiskANN. The vector dimension increases from left (128) to right (768). (b) The number of vectors that fit in a 4KB disk block for different datasets, and a node's average number of neighbors (NB) in its disk block for Starling's disk layout.}
	\label{fig:moti_dkan_vs_star}
\end{figure}


Like many disk-based systems~\cite{glist, diskgnn, atc19ssd}, to achieve good performance, disk-based ANNS must carefully design data layouts for both memory and disk to reduce the number of disk IOs and mitigate the challenges posed by disk access (e.g., long latency, low bandwidth, and 4KB block access). However, we identify two significant issues in the data layouts of DiskANN and Starling that impede efficiency.

\squishlist
\item \textit{Ineffective memory cache.} After storing the compressed vectors, DiskANN and Starling use the remaining memory as a cache (referred to as \textit{memory cache}) to keep some original vectors alongside their adjacency lists. As shown in Figure~\ref{fig:intro_mem_ratio}, when using a larger memory cache, their query throughput quickly stabilizes after an initial improvement. This is undesirable, as memory is expensive and performance should have obvious improvements with increased memory allocation. The issue arises because each individual vector is large due to high dimensionality, limiting the number of exact vectors that can be held in memory and thus leading to limited reductions in disk IOs.

\item \textit{Poor data locality on disk.} While DiskANN simply stores nodes on disk in their ID order, Starling improves the disk data layout through graph reordering~\cite{rcmk}, which increases the likelihood that a node co-locates in the same 4KB disk block as its neighbors. When reading a vector via a disk block, these co-located neighbors are also fetched, allowing some disk IOs to be avoided if these neighbors are accessed later. However, Figure~\ref{fig:moti_dkan_vs_star}(a) shows that the benefit of Starling's disk layout diminishes with increasing vector dimensions. This is because a vector becomes larger at higher dimensions, and thus a 4KB disk block holds fewer vectors, reducing the number of co-located neighbors and thereby diminishing IO savings, as indicated by Figure~\ref{fig:moti_dkan_vs_star}(b). 
This issue is concerning because there is a trend to increase vector dimensions to enhance expressiveness~\cite{power}.         

\squishend

%

To address these issues, we conducted extensive profiling and analysis to understand the impact of data layout designs on the performance of disk-based ANNS (\S~\ref{sec:insight}). Our key observation is that adjacency lists are accessed more frequently than the exact vectors with memory-resident compressed vectors, yet DiskANN and Starling treat them equally for both the memory cache and disk block storage. Specifically, each hop of the graph traversal requires to fetch the adjacency list of the current node to identify the vectors for approximate distance computation. The graph traversal visits many nodes, but only a subset of their exact vectors is needed. This is because the compressed vectors provide approximate distances, and to achieve a high recall for the search results, it suffices to compute the exact distances for the top-ranking vectors in approximate distances (this process is called \textit{re-ranking}). A secondary observation is that, capped by a maximum node degree (e.g., 64), the size of an adjacency list is generally much smaller than that of an exact vector and does not increase with vector dimensionality.

Given these insights, we propose \name{}, an ANNS index with \underline{G}raph pri\underline{O}\underline{R}itized memory cache and \underline{G}raph r\underline{E}plicated disk Lay\underline{O}\underline{U}t \underline{S}torage, featuring two key designs that prioritize adjacency lists over exact vectors.  


\squishlist

\item \textit{Graph prioritized memory cache.} Instead of keeping each adjacency list alongside its exact vector in memory cache, we retain only the adjacency lists. Since adjacency lists are significantly smaller than vectors, this allows us to cache more adjacency lists in memory, resulting in greater reductions in disk IO.

\item \textit{Graph replicated disk block.} In the disk block of each node, we explicitly include the adjacency lists of its neighbors, which we refer to as \textit{neighbor packing}. Given that adjacency lists are small, we can store many neighbor adjacency lists in a disk block. This design mitigates disk I/O when these neighbors are subsequently traversed. Although this may consume more disk space because an adjacency list could be replicated in the disk blocks of its neighbors, the space overhead remains moderate (e.g., below 2x) for high-dimensional vectors by constraining the neighbor packing to a 4KB block and given the low cost of SSD. 
 
\squishend

To make \name{} work, we introduce an additional algorithm and several system designs (\S~\ref{sec:design}). First, we modify the proximity graph traversal algorithm for query processing to take two stages: \textit{search} and \textit{refinement}. The search stage bypasses disk access for exact vectors if the required adjacency lists are in memory. In the refinement stage, we re-rank the search results by 
reading the exact vectors from disk for the top-ranking candidates based on their approximate distances. 
Second, we design a complete procedure to generate the data layout for \name{} offline, which includes determining the compression ratio for compressed vectors, loading the memory cache, and writing the disk block for each vector under a disk space budget. Finally, we propose an asynchronous disk block read pipeline that prefetches the required disk blocks for future computations to hide disk access latency.

To evaluate \name{}, we conduct experiments using 4 real datasets and compare with both DiskANN and Starling (\S~\ref{sec:eval}). The results show that \name{} consistently outperforms both baselines across the datasets and recall targets, achieving higher query throughput and lower query latency. Specifically, when targeting the same recall levels (e.g., 90\% or 95\%), \name{} improves query throughput over DiskANN and Starling by 78\% and 60\% when averaged over the datasets, while also reducing the query latency by 41\% and 35\%, respectively. Moreover, \name{} effectively improves performance with increased memory cache, and ablation studies confirm the effectiveness of our designs compared to alternatives. 

To summarize, we make the following contributions.

\squishlist
\item We identify that the data layouts of existing disk-based ANNS systems hinder their performance due to ineffective memory utilization and poor disk data locality.

\item We conduct extensive profiling and analysis to understand how data layout influences the performance of disk-based ANNS, establishing the key design principle that adjacency lists should be prioritized over exact vectors.

\item Guided by this principle, we design the \name{} system, which features a memory cache that retains adjacency lists and a disk block that includes the adjacency lists of neighbors.  

\item We implement \name{} with additional algorithm and system designs and comprehensively evaluate its performance.   
\squishend

\section{Background for Disk-based ANNS}
\label{sec:background}

\begin{figure}[!t]
	\centering
	\includegraphics[width=1\linewidth]{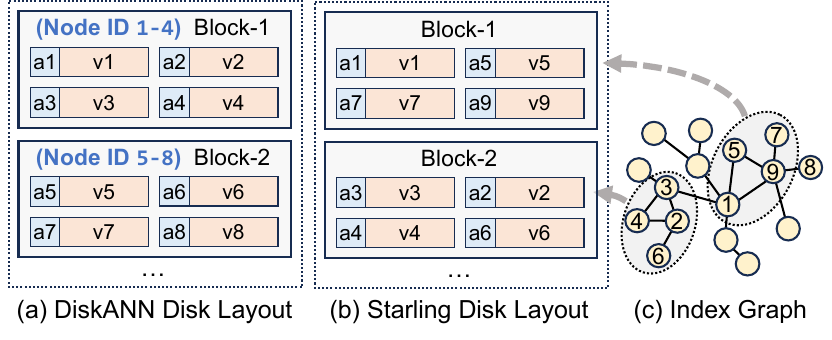}
	\caption{The disk data layout of DiskANN and Starling. (c) is a proximity graph, and `ax' and `vx' are the adjacency list and vector of node $x$, respectively. We assume that a disk block can hold 4 vectors, and as neighbors of node 1, nodes 5 and 9 reside in the same block as node 1 in (b) due to reordering.}
	\label{fig:bk}
\end{figure}


Vector search requires indexes to pinpoint a small number of vectors for each query for similarity computation. Among the indexes for high-dimensional vectors~\cite{nsg, hnsw, diskann, starling, spann, spfresh, bang, fusionanns, tree1}, proximity graph provides the best performance (i.e., in the number of similarity computations to achieve a recall target) and has many variants including HNSW~\cite{hnsw}, NSG~\cite{nsg}, and Vamana~\cite{diskann}. Both DiskANN~\cite{diskann} and Starling~\cite{starling} adopt the proximity graph index, and an illustration is provided in Figure~\ref{fig:bk}(c). Each vector is a graph node and is connected to similar vectors, and there is usually a degree limit (e.g., 32 and 64) for each node to control the number of neighbors.
Vector search is processed via a BFS-like traversal on the proximity graph. When visiting a node (called \textit{candidate}), the traversal computes the distances between its neighbors and the query and marks the node as \textit{visited}; and the nearest but unvisited node becomes the next candidate. The traversal stops when no closer nodes can be found, and the top-$k$ nearest nodes encountered in this process are returned as the search results.

DiskANN stores in memory a compressed version of all vectors to compute approximate distances with the queries, and the specific vector compression technique is product quantization (PQ)~\cite{pq}. The remaining memory serves as a \textit{node cache}, which stores the exact vectors and adjacency lists of some graph nodes. In particular, DiskANN adopts the Vamana index~\cite{diskann}, which uses the centroid of the dataset as the starting node for graph traversal (called \textit{entry node}). The node cache keeps the nodes that are within a few hops (e.g., 2-3) from the entry node since they are visited frequently. Figure~\ref{fig:bk}(a) depicts the disk data layout of DiskANN, in which the exact vector and adjacency list of a node are placed together, and each disk block contains several nodes. 


Algorithm~\ref{alg:anns} illustrates the search procedure of DiskANN. Two nearest node lists are maintained, i.e.,  $\mathcal{L}_{appr}$ is ordered by approximate distances computed using the compressed vectors, and $\mathcal{L}_{ext}$ is ordered by exact distances computed using the exact vectors.
In each iteration, the nearest unvisited node $u$ in $\mathcal{L}_{appr}$ is selected as the candidate. If $u$ is not in the node cache $NC$, the disk block containing $u$'s exact vector and adjacency list will be read from the disk. Then, the approximate distances of $u$'s neighbors are calculated and used to update $\mathcal{L}_{appr}$. The exact distance between $u$ and the query is added to $\mathcal{L}_{ext}$. After each iteration, $\mathcal{L}_{appr}$ is adjusted to retain only the top-$\mathcal{D}$ nearest nodes, where $\mathcal{D}$ is called the \textit{queue size} and usually needs to be much larger than $k$ to reach a high recall. The search terminates when all the nodes in $\mathcal{L}_{appr}$ are visited, and top-$k$ nodes in $\mathcal{L}_{ext}$ are returned as the search results.

\begin{algorithm}[t]
    \SetInd{.25em}{1em}
    \SetKwData{worker}{$\mathcal{W}$}
    \SetKwFunction{Search}{ANNS}
    \SetKwData{query}{$q$}
    \SetKwData{nb}{$v$}
    \SetKwData{node}{$u$}
    \SetKwData{entry}{$EP$}
    \SetKwData{nodecache}{$NC$}
    \SetKwData{approxList}{$\mathcal{L}_{appr}$}
    \SetKwData{fullList}{$\mathcal{L}_{ext}$}
    \SetKwProg{Fnv}{def}{:}{}

    \fullList: Nearest node list (sorted by exact distances). \\
    \approxList: Nearest node list (sorted by approximate distances). \\
    $\mathcal{D}$: The maximum length for \approxList. \\

    \BlankLine

    \Fnv{\Search{\textnormal{\texttt{\query, $k$}}}}{
        \approxList append \{\texttt{\textnormal{entry node $e$}, ApproxDist($e$, \query)}\} \\
        \While{\textnormal{\approxList have unvisited nodes}}{
            \node $\leftarrow$ the nearest unvisited node in \approxList \\
            \If{\textnormal{node \node not in node cache \nodecache}} {
                Load exact vector and adj. for  \node  from disk \\
            }
            \fullList append \{\texttt{\node, ExactDist(\node, \query)}\} \\
            \For{\textnormal{node \nb in \node's adj. list}} {
                \approxList append \{\texttt{\nb, ApproxDist(\nb, \query)}\} \\
            }
            Sort \approxList and keep the top-$\mathcal{D}$ nearest nodes \\
        }
        Sort \fullList and keep the top-$k$ nearest nodes \\
        \Return \fullList \\
    }
    
    \caption{The search procedure of DiskANN.}
    \label{alg:anns}
\end{algorithm}

Starling also keeps the compressed vectors in memory but introduces two optimizations for the data layout of DiskANN. First, for the memory cache, Starling samples a portion (e.g., 10\%) of the vectors and builds a proximity graph for these vectors (called \textit{navigation index}). A query first searches the memory resident navigation index, and the results are used as the starting points for searching the proximity graph of the entire dataset. 
Second, Starling conducts graph reordering and tries to place each node and its neighbors in the same disk block~\cite{rcmk}, as illustrated in Figure~\ref{fig:bk}(b). In particular, graph reordering assigns new IDs to the graph nodes such that nodes with similar neighbors have adjacent IDs, and Starling stores the nodes according to their new ID order. During vector search, when a disk block is loaded, Starling computes exact distances for all vectors in the block, for a portion of these vectors (e.g., 30\% with the smallest distances), neighbors are checked by computing approximate distance (called \textit{block search}). In summary, the navigation index finds good starting points while the block search may access multiple required nodes via a 4KB disk block.


SPANN~\cite{spann} adopts the IVF index~\cite{ivf} for disk-based vector search. The vectors are grouped into clusters and stored on disk, while the cluster centers are kept in memory. To conduct vector search for a query, SPANN first finds the nearest cluster centers in memory and then loads the corresponding clusters from the disk for exact distance computation. Recent research~\cite {starling} shows that Starling outperforms SPANN because the IVF index requires to access many more vectors than proximity graph to reach the same recall.

\begin{table}[!t]
    \caption{Statistics of the datasets. \textbf{Tot.N} is the number of vectors, \textbf{Dim.} is vector dimension, \textbf{Metric} is the measure for vector similiarty, and \textbf{Tar.} is the target recall@10 for search.}
    \setlength{\tabcolsep}{3pt}
    \label{tab:datasets}
    \centering
    \fontsize{9}{11}\selectfont
    \begin{tabular}{c|cccccc}
        \toprule[1pt]
        \textbf{Name} & \textbf{Tot. N} & \textbf{Dim.} & \textbf{Type} & \textbf{Metric} & \textbf{Size (GB)} & \textbf{Tar.} \\
        \midrule
        Sift~\cite{bigann} & 100M & 128 & uint8 & L2 & 11.9 & 95\%\\
        Deep~\cite{text2image} & 100M & 96 & float32 & L2 & 35.8 & 95\% \\
        Wiki~\cite{wiki} & 100M & 384 & float32 & L2 & 143.3 & 95\% \\
        Text2Image~\cite{text2image} & 100M & 200 & float32 & IP & 74.5 & 90\% \\
        Laion-T2I~\cite{laion} & 100M & 512 & float32 & Cosine & 191.7 & 90\% \\
        Laion-I2I~\cite{laion5b} & 100M & 768 & float32 & Cosine & 286.4 & 95\% \\
        \bottomrule[1pt]
    \end{tabular}
\end{table}

\section{Observations and Insights}
\label{sec:insight}

In this part, we conduct extensive profiling and analysis to understand the influence of data layout on the performance of disk-based ANNS. Table~\ref{tab:datasets} reports the statistics of the utilized datasets, and the detailed experiment settings can be found in \S~\ref{subsec:settings}. Note that the query throughput and disk IOs are measured when reaching the target recalls in Table~\ref{tab:datasets}. To our knowledge, this is the first comprehensive profiling for data layout decisions, and we also draw insights to guide data layout designs and conduct in-depth analysis.




\subsection{Accuracy of the Compressed Vectors}
\label{subsec:analysis pq}



The first data layout decision is the compression ratio of the compressed vectors that are kept in memory. Although there are many vector quantization techniques that work in different ways~\cite{pq, opq, rabitq, lvq}, a higher compression ratio always makes the compressed vectors smaller, while the approximate distance becomes cheaper to compute but less accurate.

\begin{figure}[!t]
	\centering
	\includegraphics[width=1\linewidth]{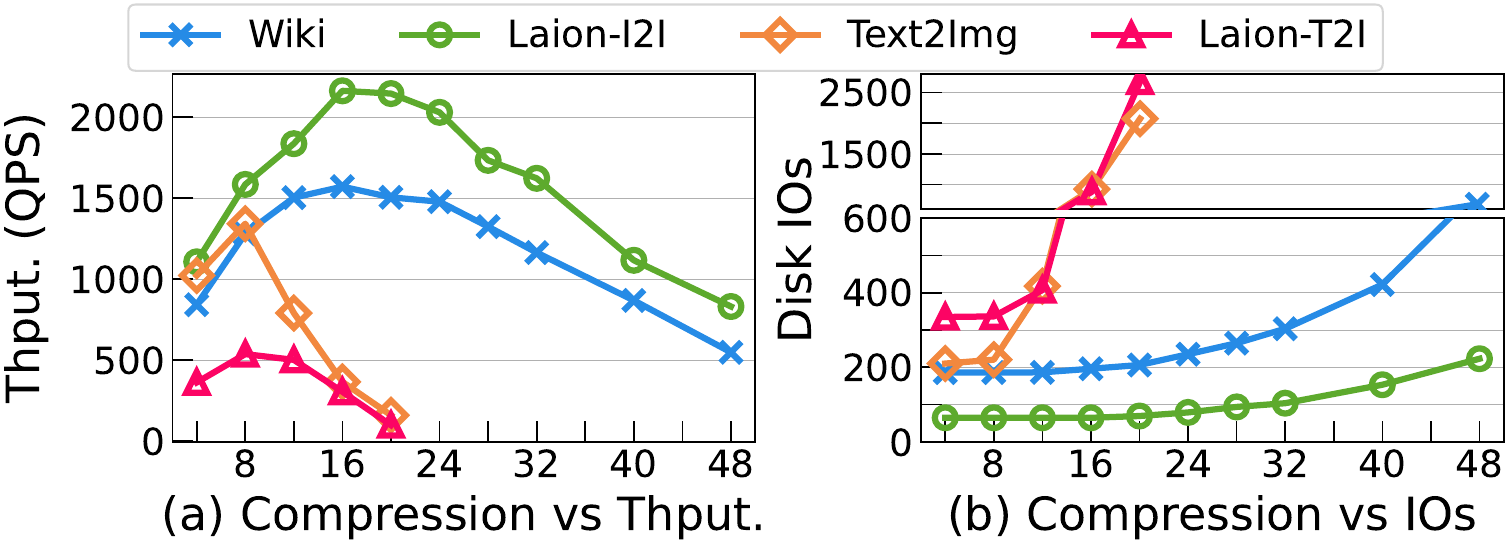}
	\caption{DiskANN's query throughput and per query disk IOs when using different compression ratios for the memory-resident compressed vectors. We disable the memory cache for this experiment, and the results are similar for Starling.}
	\label{fig:analysis_pq}
\end{figure}

As shown in Figure~\ref{fig:analysis_pq}(b), the disk IOs for each query are stable initially but increase quickly after the compression ratio goes beyond a threshold. This is because at low compression ratios, the approximate distances are accurate and can identify good candidates (i.e., those with small exact distances to the query) for the graph traversal. As such, to reach the same recall, the length of the search path (i.e., the queue size $\mathcal{D}$) and hence the number of disk accesses resemble using uncompressed exact vectors. At high compression ratios, the approximate distances become inaccurate and lead the graph traversal to visit vectors that are dissimilar to the query, which prolongs the search path and blows up disk IOs. Figure~\ref{fig:analysis_pq}(a) shows that query throughput, the main performance metric to optimize, first increases and then decreases with the compression ratio, making one compression ratio the optimal. This is because at smaller compression ratios, approximate distance computation becomes more expensive, while more disk IOs are required at higher compression ratios. In fact, the optimal is the highest compression ratio that does not blow up disk IOs.

\vspace{0.3em}
\textbf{Insight 1.} \textit{The compression ratio of approximate vectors has an optimal, and smaller or larger compression ratio degrades performance.}

Another observation is that the optimal compression ratios are different for the datasets, and cross-modal datasets (e.g., Text2Image and Laion-T2I) have smaller optimal than single-modal datasets (e.g., Wiki and Laion-I2I). In particular, cross-modal datasets use queries from one modality to search objects from another modality, for instance, Text2Image uses text embeddings to search image embeddings; they require more accurate compressed vectors because the distance gap between the vectors similar and dissimilar to a query is smaller compared with the single-modal datasets~\cite{roargraph}.

\subsection{Demand for the Exact Vectors}


%

Since the compressed vectors provide approximate distances with reasonable accuracy, we wonder whether it is necessary to compute exact distances for all the candidate vectors as in Algorithm~\ref{alg:anns} (i.e., for $\mathcal{L}_{ext}$). 
Instead, we may only compute exact distances for the top-ranking vectors in $\mathcal{L}_{appr}$ (see Algorithm~\ref{alg:anns}) because $\mathcal{L}_{appr}$  records the approximate distances of the encountered vectors, and the real top-$k$ neighbors of a query should have a good probability to rank high in $\mathcal{L}_{appr}$. To implement this idea, we define a \textit{refinement ratio} $\sigma$ and compute exact distances for $\mathcal{D}_{r}=\sigma \mathcal{D}$ vectors, where $\mathcal{D}$ is the size of $\mathcal{L}_{appr}$  (i.e., \textit{queue size}). We also modify DiskANN to compute exact distances after the graph traversal terminates.

\begin{figure}[!t]
	\centering
	\includegraphics[width=1\linewidth]{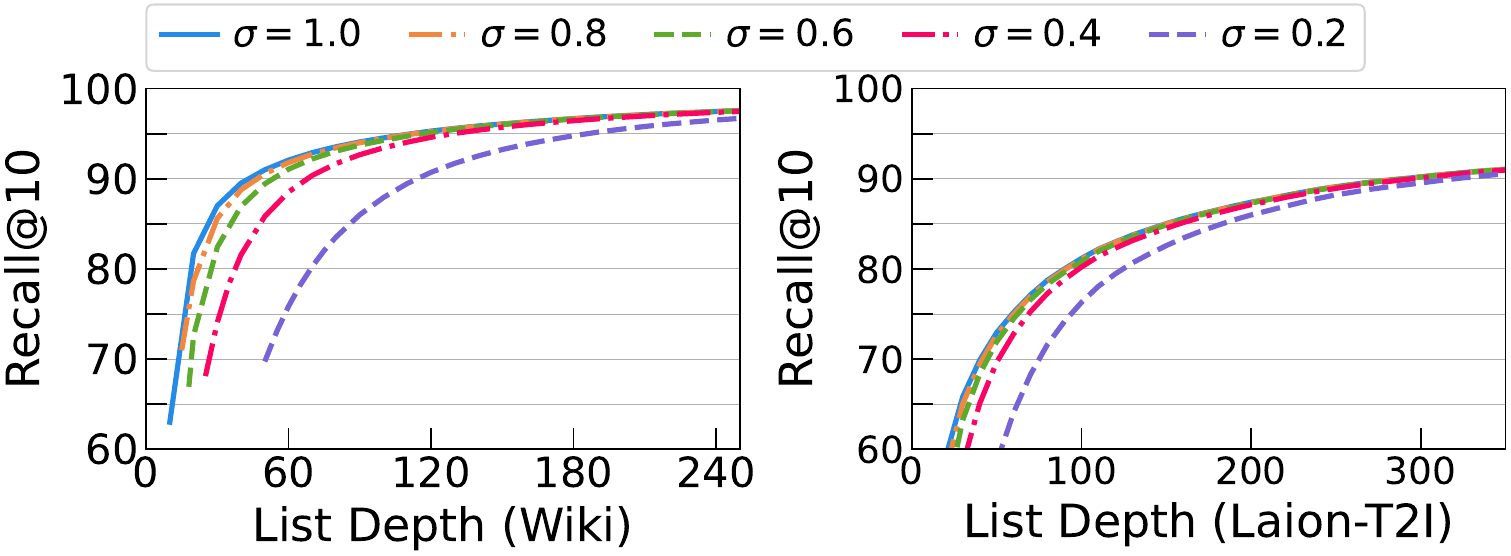}
	\caption{Recall@10 for different refinement ratio $\sigma$ and queue size $\mathcal{D}$. Tested on the Wiki and Laion-T2I datasets. We disable the memory cache for this experiment.}
	\label{fig:sigma}
\end{figure}

Figure~\ref{fig:sigma} shows the impact of the refinement ratio $\sigma$ and queue size $\mathcal{D}$ on the quality of the search results. We observe that to achieve a high recall (e.g., 95\% for Wiki and 90\% for Laion-T2I), $\mathcal{D}$ needs to be 10-20$\times$ of the target top-$k$; and at large $\mathcal{D}$, using a moderate $\sigma$ (e.g., 0.5) does not harm result quality. Since the number of candidate vectors visited by the graph traversal is no smaller than $\mathcal{D}$\footnote{Algorithm~\ref{alg:anns} terminates when $\mathcal{L}_{appr}$  has no unvisited candidates, and the size of $\mathcal{L}_{appr}$ is $\mathcal{D}$. As such, the graph traversal visits at least $\mathcal{D}$ candidates.}, this allows to reduce a considerable portion of the exact distance computations.

\vspace{0.3em}
\textbf{Insight 2.} \textit{Re-ranking the vectors that rank top in approximate distance suffices to achieve a high recall.}

\subsection{Contents of the Memory Cache}
\label{subsec:memory cache}


The insight above implies that the adjacency lists are more important than the exact vectors. That is, when the graph traversal checks each candidate vector, its adjacency list is required to identify the neighbors for approximate distance computation, while its exact vector is not required if its approximate distance cannot rank top-$\mathcal{D}_{r}$ in the final $\mathcal{L}_{appr}$. This observation suggests that the memory cache should store only the adjacency lists rather than both exact vectors and adjacency lists as in DiskANN and Starling. To make the idea work, we modify Algorithm~\ref{alg:anns} to skip disk access if the required adjacency list is in memory and load the exact vectors for re-ranking after the graph traversal terminates (see the detailed algorithm in \S ~\ref{subsec:decoupling}).



\begin{figure}[!t]
	\centering
	\includegraphics[width=1\linewidth]{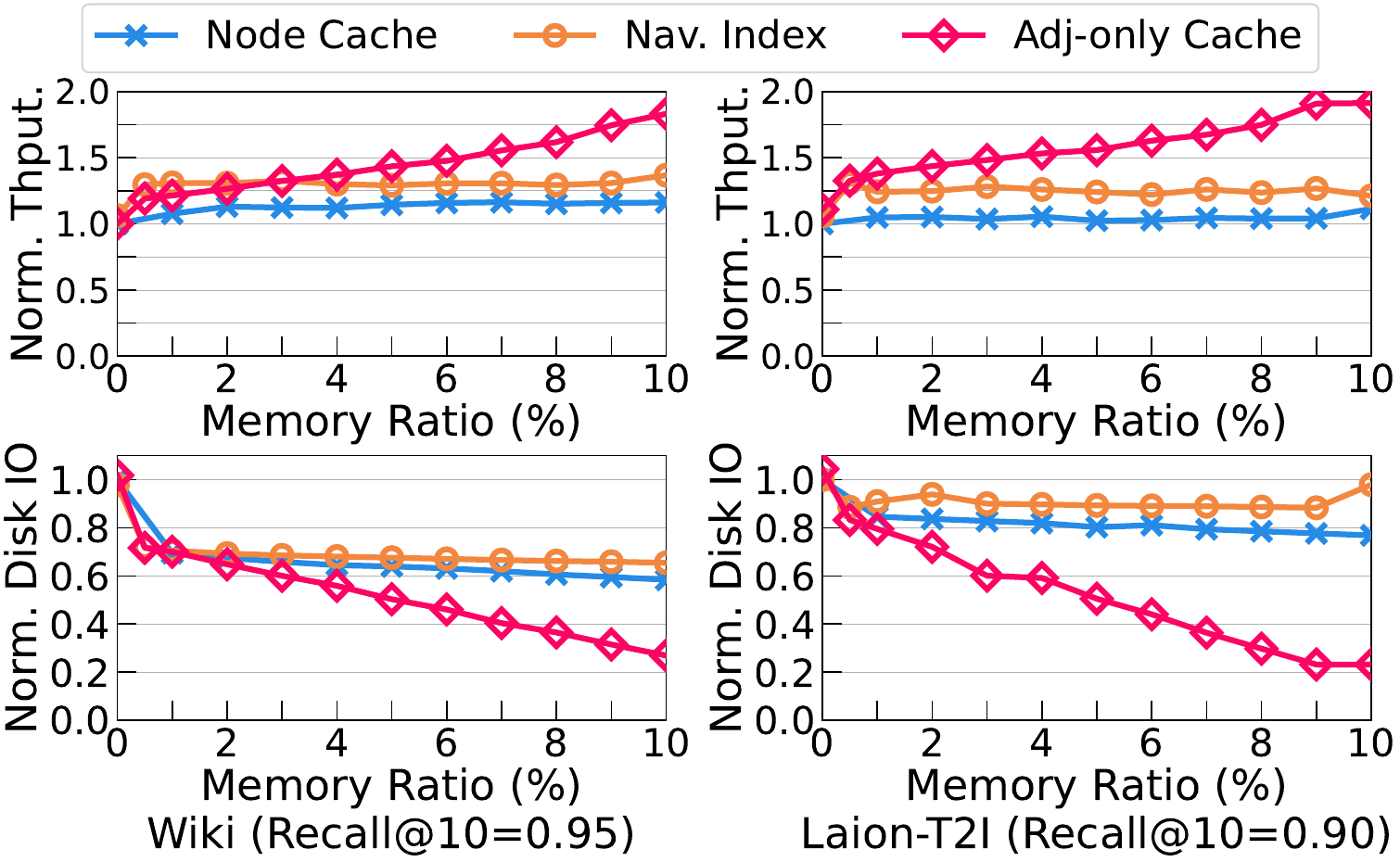}
	\caption{The throughput and disk IOs for different memory caches when varying the size of the memory cache. Normalized by the case that does not use memory cache.}
	\label{fig:moti_graph_only}
\end{figure}

Figure~\ref{fig:moti_graph_only} compares the query throughput and disk IOs of DiskANN's node cache, Starling's navigation index, and our adjacency-only cache. The results echo Figure~\ref{fig:intro_mem_ratio}, i.e., DiskANN and Starling only observe marginal IO reduction and thus throughput improvement when increasing the memory size beyond a small threshold. This is because after including a small number of frequently visited vectors (e.g., those close to the entry node), the remaining vectors are accessed more uniformly; the cache size is small compared to the entire dataset, and thus the IO reduction is not obvious. In contrast, our adjacency-only cache achieves significant query throughput improvement and disk IO reduction when increasing the memory size. This validates our conjecture, and the following analysis explains why the adjacency-only cache works.    

\stitle{Analysis} Holding the adjacency lists in the memory cache outperforms holding both adjacency lists and vectors when
\begin{equation}\label{equ:memory cache}
\frac{C}{N(S_v+S_a)}< \frac{C(1-\sigma)}{NS_a}\ \Rightarrow \ S_a<\frac{1-\sigma}{\sigma}S_v,
\end{equation}
where $C$ is the size of the memory cache, $N$ is the number of vectors in the dataset, $S_v$ and $S_a$ are the sizes of an exact vector and an adjacency list, respectively, and $\sigma\approx0.5$ is the refinement ratio defined previously. Assume that the graph traversal visits all nodes with an equal probability\footnote{If the visit probabilities are different for the nodes, we can select the popular nodes for both the adjacency-only cache and adjacency-vector cache.}, the adjacency-vector coupled cache holds $C/(S_v+S_a)$ vectors, and thus its ratio of IO reduction is $C/N(S_v+S_a)$. The adjacency-only cache holds $C/S_a$ adjacency lists and reduces the disk accesses for adjacency lists by a ratio of $\beta=C/(NS_a)$; without the memory cache, the ANNS search will conduct $\mathcal{D}$ disk accesses; with the cache, the total disk accesses $\mathcal{T}_a$ and the access reduction ratio $\mathcal{A}_r$ becomes
\begin{equation}\label{eq:reduction}
\mathcal{T}_a=\mathcal{D}(1-\beta)+\sigma \mathcal{D} \beta,\ \mathcal{A}_r=\frac{\mathcal{D}-\mathcal{T}_a}{\mathcal{D}}=\beta(1-\sigma).
\end{equation}
Note that in Eq~\eqref{eq:reduction}, we compute the disk IO reduction w.r.t. the $\mathcal{D}$ accesses for the adjacency-vector cache to align both cases.
As such, Eq~\eqref{equ:memory cache} compares the IO reduction ratios of the two cache strategies, and a larger reduction is better.


Eq~\eqref{equ:memory cache} easily holds because an adjacency list records tens of neighbor IDs, while a vector can have hundreds or even thousands of dimensions. Take the Wiki dataset for an example, we have $S_v=1536 \text{B}$ and $S_a=200 \text{B}$. When using a memory cache that is 10\% of the dataset size, the adjacency-only cache holds the adjacency lists for 88\% of the nodes, almost avoiding disk read during graph traversal, while the adjacency-vector coupled cache holds only 10\% of the nodes.

\vspace{0.3em}
\textbf{Insight 3.} \textit{Caching the adjacency lists in memory is more effective than the exact vectors.}

\begin{figure}[!t]
	\centering
	\includegraphics[width=0.9\linewidth]{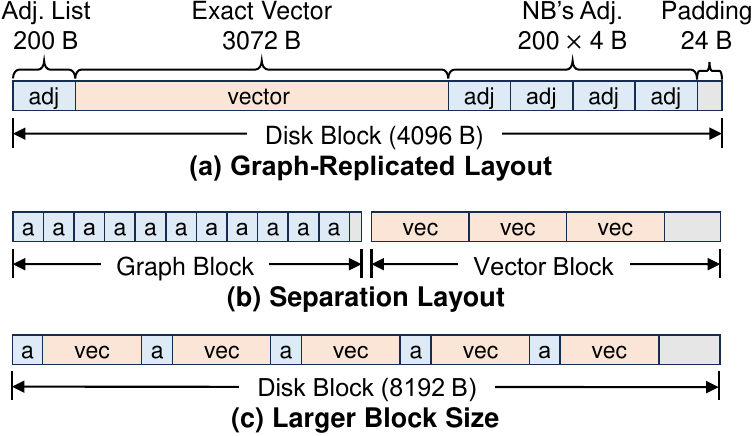}
	\caption{Disk data layouts. (a) an 4KB illustration of our graph-replicated layout on the Laion-I2I dataset. (b) separation layout, using separate graph and vector blocks; (c) using larger blocks (8KB) to place more nodes in a block.}
	\label{fig:layout}
\end{figure}

\subsection{Format of the Disk Block}
\label{subsec:disk block}

Inspired by the importance of the adjacency lists, we introduce a disk block format that explicitly stores the adjacency lists of a node's neighbors along with the node. An illustrative example is provided in Figure~\ref{fig:layout}(a) by assuming 4KB disk block size, and after keeping the exact vector and adjacency list of a vector, the adjacency lists of some neighbors are included to fill a disk block. We call it the \textit{graph-replicated layout} since an adjacency list may be stored multiple times in the disk blocks of its neighbors. It consumes more disk space but we can control the space blow up as will be described in \S ~\ref{subsec:storagelayout}, and disk capacity is cheap. The layout allows to read the neighbor adjacency lists along with the vector, and when these neighbors become the candidates, disk IOs can be skipped for their adjacency lists. \modify{Note that the disk block size is not constrained to 4KB, and \name{} can use larger block sizes (e.g., 8KB and 16KB) for vectors with very high dimension (e.g., 1536 or 3072) like DiskANN. In these cases, the neighbor adjacency lists can also fill in the remaining space like for 4KB disk block.}

\begin{figure}[!t]
	\centering
	\includegraphics[width=1\linewidth]{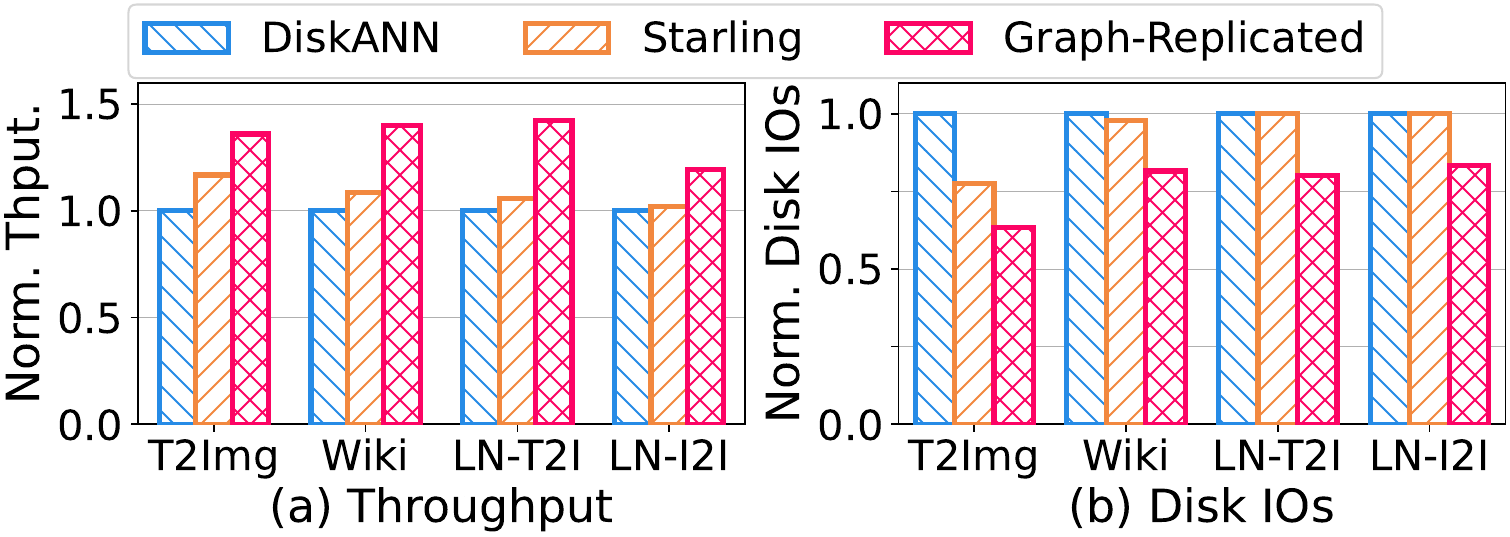}
	\caption{Normalized throughput and disk IOs for DiskANN's, Starling's, and our graph-replicated disk data layout.}
	\label{fig:moti_gr}
\end{figure}

Figure~\ref{fig:moti_gr} compares the performance of our graph-replicated disk layout, Starling's disk layout, and DiskANN's disk layout with the other optimizations disabled (e.g., node cache and navigation index). The result shows that the graph-replicated layout consistently outperforms the other layouts for all 4 datasets. Figure~\ref{fig:moti_gr}(b) also explains why Starling has no improvement over DiskANN for the Laion-I2I dataset in Figure~\ref{fig:moti_dkan_vs_star}. That is, each vector in Laion-I2I takes around 3KB, and thus Starling can only store 1 node in the same 4KB disk page, yielding no IO reduction and throughput improvement. In contrast, since each adjacency list is small, our graph-replicated disk layout can still keep several neighbor adjacency lists along with the vector in a 4KB disk page.


\stitle{Analysis} Our graph-replicated disk block outperforms Starling's disk block when
\begin{equation}\label{equ:disk page}
\theta\frac{B'}{S_v+S_a}< (1-\sigma) \theta \frac{B'}{S_a}\ \Rightarrow\  S_a<\frac{1-\sigma}{\sigma}S_v,
\end{equation}
which easily holds since $\sigma \approx 0.5$,  making Eq~\eqref{equ:disk page} essentially $S_a<S_v$. We consider the number of future disk accesses that can be avoided after reading a disk block in the two different formats. Let $B'=B-S_v-S_a$, where $B$ is the block size, $B'/(S_v+S_a)$ is the maximum number of neighbors Starling can hold in a disk block. Assuming each of these neighbors has a probability of $\theta$ to become the candidate in the future, Starling reduces at most $\theta B'/(S_v+S_a)$ disk accesses.  For the graph-replicated disk block, it can store the adjacency lists of $ B'/S_a$ neighbors and thus avoids $\theta B'/S_a$ disk accesses for the adjacency lists during graph traversal; for re-ranking, it needs  $\sigma \theta B'/S_a$ additional disk accesses for the exact vectors of these skipped adjacency lists; and thus the total disk access reduction is $( 1-\sigma)\theta B'/S_a$.\footnote{We ignore integer rounding in the analysis for simplicity, e.g., the graph-replicated disk block actually holds $\lfloor  B'/S_a \rfloor$ neighbor adjacency lists. We overestimate the number of co-located neighbors for Starling because its graph reordering does not ensure that a disk block only holds the neighbors of a vector, which is also illustrated in Figure~\ref{fig:moti_dkan_vs_star}(b).}


\vspace{0.3em}
\textbf{Insight 4.} \textit{For high-dimensional vector datasets, packing the adjacency lists of a node's neighbors in its disk block improves performance.}
\vspace{0.3em}

\stitlestart{Alternatives}
Besides our graph-replicated disk block, two alternative disk block formats may also improve data locality, following the same idea of packing neighbors. 
The first format uses two types of disk blocks as shown in Figure~\ref{fig:layout}(b), i.e., \textit{graph blocks} that contain only adjacency lists and are accessed during graph traversal, and \textit{vector blocks} that contain only vectors and are accessed during re-ranking~\cite{separation}. This format allows one graph block to hold more neighbor adjacency lists, which increases the chance of reducing the disk IOs for adjacency lists during graph traversal. However, a considerable portion of the nodes (i.e., $\sigma=0.5$) need one disk access for their exact vectors during re-ranking, which may outweigh the disk IO savings for the adjacency lists. The second format uses a larger block size (e.g., 8KB) to include more nodes in a block, as shown in Figure~\ref{fig:layout}(c). Although this format allows fetching more nodes via a single disk block, it wastes disk bandwidth because each disk block is larger. It has been reported that using a 4KB block size can achieve the full bandwidth of modern NVMe SSDs~\cite{vldbssd}. As such, the disk bandwidth should be used more judiciously by making access decisions individually for each 4KB block.    
In \S\ref{subsec:designchoices}, we compare our graph-depreciated disk layout with the two alternatives and show that they yield inferior performance.

\section{The Gorgeous System}
\label{sec:design}

\begin{figure}[!t]
	\centering
	\includegraphics[width=1\linewidth]{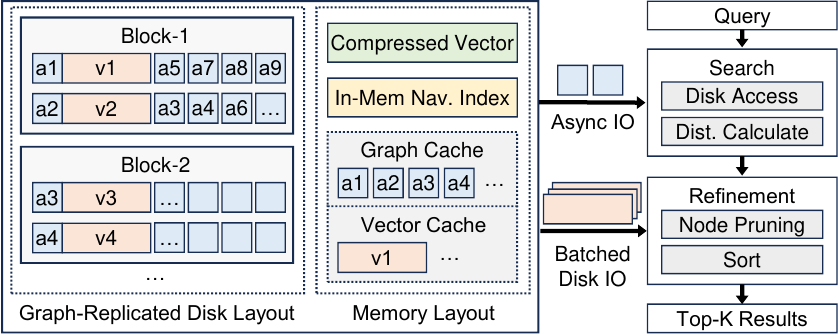}
	\caption{The storage layout and workflow of \name{}.}
	\label{fig:arch}
\end{figure}

\name{} is a general system for disk-based ANNS on a single machine. Like Starling and DiskANN, \name{} runs with a small memory (e.g., 10\% of the dataset size) and uses disk as the primary data storage. \name{} can plug in different proximity graph indexes~\cite{nsg, diskann} and vector quantization algorithms~\cite{pq, opq, rabitq} as all proximity graphs adopt the same graph traversal procedure for query processing, and vector quantization is used as a black box to compress the vectors and compute approximate distances.
Figure~\ref{fig:arch} shows the workflow and storage layout of \name{}.
When \name{} receives a query, it searches for its approximate top-$k$ nearest vectors using the data stored in both memory and disk, and each query is handled by one thread. 
In this part, we put together the insights from \S~\ref{sec:insight} and present the key designs of \name{}, which includes the graph-vector decoupled storage layout ($\S$~\ref{subsec:storagelayout}), a two-stage search algorithm that fits the decoupled storage layout ($\S$~\ref{subsec:decoupling}), and asynchronous block prefetch to hide the latency of disk access ($\S$~\ref{subsec:pipeline}).


\subsection{Graph-vector Decoupled Storage Layout}
\label{subsec:storagelayout}

DiskANN and Starling always store the vector and adjacency list of a node together for both memory and disk. Observing that the adjacency lists are more important than the vectors, \name{} decouples the storage decisions for vectors and adjacency lists and prioritizes the adjacency lists.    

\stitlestart{Graph prioritized memory cache}
After storing the compressed vectors, the remaining memory space can serve as a cache. We mainly use the cache to store the adjacency lists of the proximity graph index as analyzed in \S\ref{subsec:memory cache} (called \textit{graph cache}). However, Figure~\ref{fig:moti_graph_only} shows that keeping a small navigation index in memory also improves performance, and thus we also sample a small portion of the vectors (e.g., 0.5\%) to build the navigation index. To select the nodes (whose adjacency lists are kept) in the graph cache, 
we order the nodes by their minimum distances to the nodes in the navigation index and cache the nodes with the smallest distances. The rationale is that nodes of the navigation index serve as entry nodes for graph traversal, and nodes close to these entry nodes have higher probabilities of being visited.  


\stitle{Graph replicated disk block} \name{} uses the graph-replicated disk block format in Figure~\ref{fig:layout}(a) to improve data locality. To control the disk space blow up caused by replicating the adjacency lists, we allow users to specify the number of neighbor adjacency lists that are packed for each node (denoted as $R$). 
Compared to Starling, the disk space blow up of \name{} is $((1+R)S_a+S_v)/(S_a+S_v)$\footnote{We ignore integer rounding in the calculation for simplicity, e.g., the block capacity of Starling and \name{} should be $\lfloor  B/(S_a+S_v) \rfloor$ and $\lfloor  B/((1+R)S_a+S_v) \rfloor$, respectively. $B$ is the block size.}, since each node requires an extra storage of $RS_a$ bytes for the packed adjacency lists. As discussed in \S\ref{subsec:disk block}, we constrain that the disk block of each vector takes up at most one 4KB disk page.

Given a proximity graph index and $R$, for each node $u$, our graph-replicated disk layout selects the $R$ neighbors with the smallest exact distances to $u$ as the neighbors whose adjacency lists are packed with $u$. This is because the close neighbors of a node are more likely to be visited after the node. To balance the replication of different adjacency lists, we constrain that each adjacency list can be replicated at most $R+1$ times. This reduces the chance that different nodes pack the same adjacency list, which wastes disk bandwidth during query processing. To implement the constraint, we conduct neighbor packing for the nodes one by one and avoid packing nodes whose adjacency lists have been packed for $R+1$ times. We also store the IDs of the packed neighbors on disk such that their exact vectors can be fetched during re-ranking. After generating the disk blocks for the nodes, we store the disk blocks in the order of their node IDs.

With our graph replicated format, a disk page may still contain more than one node (i.e., when the vector dimension and $R$ are small). In this case, Starling's graph reordering can place neighbors on the same disk page for better locality. We do not adopt this optimization because its gain is marginal with our graph replicated format, and additional memory is required to store the mapping from node IDs to disk pages.

\stitle{Memory cache planning}
\name{} determines the data structures to store in the memory and their sizes as follows:

\squishlist
\item[\ding{182}]\textit{Determine the compression ratio.}
For the compressed vector, \name{} searches the optimal compression ratio (i.e., as in \S~\ref{subsec:analysis pq}) that fits in memory using a small sampled dataset $\mathcal{S}'$ (e.g., 1\% of the complete dataset) and query set $\mathcal{Q}'$. \name{} uses PQ~\cite{pq} by default for vector compression and tries different compression ratios on $\mathcal{S}'$  by testing the performance of  $\mathcal{Q}'$. We observe that the optimal compression ratios are the same for the complete and sampled datasets, and a small sampled dataset allows to enumerate the compression ratios much faster due to faster compression. This is because the accuracy of the compressed vectors is more related to the relative error than the dataset size~\cite{rabitq}.

\item[\ding{183}]\textit{Determine the navigation index.} 
\modify{\name{} randomly samples a small portion of the vectors in the dataset and uses them to construct the in-memory navigation index. This is because a larger navigation index does not improve performance as shown in  Figure~\ref{fig:intro_mem_ratio}(b). The reason is that with typical graph degrees (e.g., 64, 128), the ground-truth neighbors of queries are already within 1-2 hops of the entry points identified by a small navigation index (e.g., 0.5\%).} We profile whether the navigation index improves performance by running the sampled query set $\mathcal{Q}'$. If it does not improve performance, the navigation index is disabled. One such example is the Text2Image dataset in Figure~\ref{fig:intro_mem_ratio}(b). 

\item[\ding{184}]\textit{Determine the graph cache and node cache.} 
For the rest of the memory space, \name{} stores the adjacency lists. If there is still remaining space after storing the adjacency lists of all nodes, we store exact vectors and follow the same rule as the graph cache to select the vectors to store.

\squishend

As shown in the middle plot of Figure~\ref{fig:arch}, \name{} may hold compressed vectors, navigation index, graph cache, and node cache in memory. Memory cache planning is fast for \name{} and takes less than 5\% of the overall index construction time, which is dominated by building the proximity graph.

\stitle{\modify{Discussions}} \modify{Our memory cache planning relies on query-based parameter tuning. This is practical since applications of vector search can easily collect some sample queries, and many other parameters for vector search (e.g., the degree limit for proximity graph and queue size for graph traversal) are also determined by query-based tuning~\cite{hnsw,nsg}. For a fair performance evaluation, we exclude the tuning queries from performance testing. For applications that need to update the vectors (i.e., with insertions and deletions), \name{} can adopt existing algorithms  (e.g., FreshDiskANN~\cite{freshdiskann}) to update the proximity graph index in a streaming manner (i.e., without full index rebuilding). As our disk layout replicates adjacency lists, a graph update needs to change several disk pages. To reduce the update cost, we can  commit the graph updates to the disk layout periodically, with the advantage of batching multiple updates to the same disk block. More delicate designs are possible (e.g., cache updates and install updates when the data is fetch during query processing) but we leave them for future work.}

\subsection{Two-stage Search Algorithm}
\label{subsec:decoupling}

\begin{algorithm}[t]
    \SetInd{.25em}{1em}
    \setlength{\floatsep}{-2pt} 
    \setlength{\textfloatsep}{0pt}
    \setlength{\intextsep}{0pt}
    \SetAlgoInsideSkip{-2pt}
    \SetKwData{worker}{$\mathcal{W}$}
    \SetKwFunction{Search}{ANNS}
    \SetKwFunction{Expand}{Expand}
    \SetKwData{query}{$q$}
    \SetKwData{nb}{$v$}
    \SetKwData{node}{$u$}
    \SetKwData{entry}{$EP$}
    \SetKwData{graphcache}{$GC$}
    \SetKwData{vectorcache}{$VC$}
    \SetKwData{approxList}{$\mathcal{L}_{appr}$}
    \SetKwData{fullList}{$\mathcal{L}_{ext}$}
    \SetKwProg{Fnv}{def}{:}{}

    \fullList: Nearest node list (sorted by exact distances). \\
    \approxList: Nearest node list (sorted by approx. distances). \\
    $\mathcal{D}$: The maximum length for \approxList. \\
    $\mathcal{D}_{r}$: The number of vectors to re-rank, $\mathcal{D}_{r}=\sigma \mathcal{D}$. \\

    \BlankLine

    \Fnv{\Expand{\textnormal{\texttt{\query, \node}}}}{
        \For{\textnormal{node \nb in \node's adj. list}} {
            \approxList append \{\texttt{\nb, ApproxDist(\nb, \query)}\} \\
        }
        Sort \approxList, keep the top-$\mathcal{D}$ nodes \\
    }

    \BlankLine
    \Fnv{\Search{\textnormal{\texttt{\query, $k$}}}}{
        \approxList append \{\texttt{\textnormal{entry $e$}, ApproxDist($e$, \query)}\} \\ 
        \While{\textnormal{\approxList have unvisited nodes}}{
            \node $\leftarrow$ the nearest unvisited node in \approxList \\
            \If{\textnormal{\node in graph cache \graphcache}} {
                \texttt{Expand(\query, \node)} \\
            }
            \Else {
                Load exact vector and adj. list for \node \\
                \fullList append \{\texttt{\node, ExactDist(\node, \query)}\} \\
                \texttt{Expand(\query, \node)} \\
                \For{\textnormal{node \nb in \node's block\ \textbf{and} \nb in \approxList}} {
                    \texttt{Expand(\query, \nb)} \\
                }
            }
        }
        Keep the top-$\mathcal{D}_{r}$ nodes in \approxList \\ 
        \For{\textnormal{node \node in \approxList\ \textbf{and} \node not in \fullList}} {
            \If{\textnormal{\node not in vector cache \vectorcache}} {
                Load exact vector for \node \\
            }
            \fullList append \{\texttt{\node, ExactDist(\node, \query)}\} \\
        }
        Sort \fullList, keep the top-$k$ nodes \\
        \Return \fullList \\
    }

    \caption{Two-stage ANNS algorithm for query processing using the decoupled storage layout.}
    \label{alg:anns sep}
\end{algorithm}


To adapt to the decoupled storage layout, \name{} utilizes Algorithm~\ref{alg:anns sep} to conduct ANNS, which consists of a \textit{search stage} (lines 10-20) and a \textit{refinement stage} (lines 21-26). The search stage mainly conducts graph traversal by computing approximate distances: if the candidate node $u$ is in the memory resident graph cache, disk access is bypassed, and \name{} computes the approximate distances between its neighbors and the query by reading $u$'s adjacency list from the graph cache (lines 13-14). 
Otherwise, \name{} loads the block of $u$ from disk, which includes $u$'s exact vector, adjacency list, and the adjacency lists of some neighbors of $u$. \name{} computes exact distance for $u$ (line 17) and approximate distances for $u$'s neighbors (line 18). A subtle point is that \name{} also checks for each neighbor of $u$ whose adjacency list is fetched (denoted as $v$), whether it belongs to $\mathcal{L}_{appr}$ (line 19). If this holds, $v$ has a small approximate distance to the query and may be visited later; as such, \name{} computes approximate distances for $v$'s neighbors using its adjacency list (line 20).

After the search stage, $\mathcal{L}_{appr}$ records the top-ranking vectors in approximate distance, and $\mathcal{L}_{ext}$ keeps the exact distances computed using the loaded disk blocks. The refinement stage computes exact distances for the top-$\mathcal{D}_{r}$ nodes in $\mathcal{L}_{appr}$ to determine the final research results. This is because the real  top-$k$ neighbors of the query have a high probability to rank high in $\mathcal{L}_{appr}$ since the approximate distances have reasonable accuracy. By default, we set the refinement ratio $\sigma=0.5$ and $\mathcal{D}_{r}=\sigma \mathcal{D}$, which provides good accuracy as shown in Figure~\ref{fig:sigma}. To compute exact distance for each of the top-$\mathcal{D}_{r}$ nodes, \name{} first check if it is already in $\mathcal{L}_{ext}$, then checks if it is in the vector cache, and loads the vector from disk otherwise (lines 22-25). Finally, \name{} returns the top-$k$ nodes in exact distance as the ANNS results (line 27).

\subsection{Asynchronous Block Prefetch}
\label{subsec:pipeline}


DiskANN uses synchronous IO to read the disk blocks, and computation is halted until the IO finishes. Starling improves DiskANN by checking the non-candidate nodes in the loaded disk blocks while waiting for the IOs of new disk blocks to complete. Since each disk block contains only a few nodes (i.e., 1-3) for high-dimensional vectors and the computation is lightweight for checking non-candidate nodes, the query processing thread may still be idle waiting for disk access, which prolongs query latency. 

\begin{figure}[!t]
	\centering
	\includegraphics[width=0.9\linewidth]{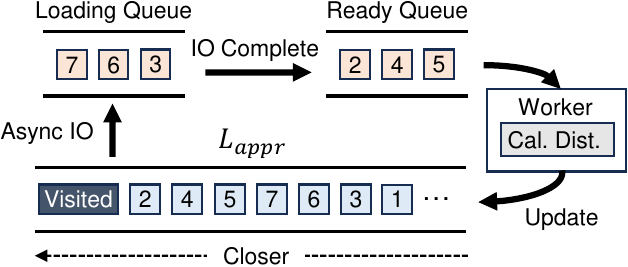}
	\caption{The workflow of asynchronous block prefetch.}
	\label{fig:pipeline}
\end{figure}

To avoid stalling computation and reduce query latency, \name{} adopts the asynchronous disk block prefetch pipeline in Figure~\ref{fig:pipeline}. In particular, we use two queues for the IO requests, i.e., \textit{loading queue} for the disk blocks whose IOs are in-flight, and \textit{ready queue} for the disk blocks that have already been fetched to memory. Like DiskANN and Starling, \name{} issues IO requests for a batch of top-ranking but unvisited candidate nodes in $\mathcal{L}_{appr}$ each time, and the batch size is called the \textit{beam width} for graph traversal. Once the disk block of a candidate node is loaded, \name{} moves it from the loading queue to the ready queue, and the thread handling the query keeps getting data from the ready queue for computation and updating $\mathcal{L}_{appr}$. Our pipeline is implemented using the asynchronous interface of \texttt{libaio}~\cite{aio} and ensures that computation will not be blocked as long as the ready queue has disk blocks.

\stitle{Other optimizations}
In the refinement stage, as the disk accesses for the exact vectors have no dependencies, their IO requests are submitted in batches to fully utilize multiple disk channels and reduce latency. Disk IO libraries (e.g., \texttt{libaio}~\cite{aio} and \texttt{io\_uring}~\cite{iouring}) provide event queues and allow obtaining the IO results one by one. \name{} utilizes the event queue and computes the exact distance for each vector once its disk IO finishes, instead of waiting for the entire IO batch.

\section{Experimental Evaluation}
\label{sec:eval}

We extensively evaluate \name{} and compare with state-of-the-art systems for disk-based ANNS. The key findings include:

\squishlist
\item \name{} consistently outperforms the baselines, achieving significantly higher query throughput and lower query latency.
\item Our two key designs, graph prioritized memory cache and graph replicated disk block, are effective in reducing the disk IOs and outperform alternative designs.
\item Other designs of \name{} (e.g., asynchronous block prefetch) also enhance system performance.
\squishend

\begin{figure*}[!t]
    \centering
    \includegraphics[width=2.1\columnwidth]{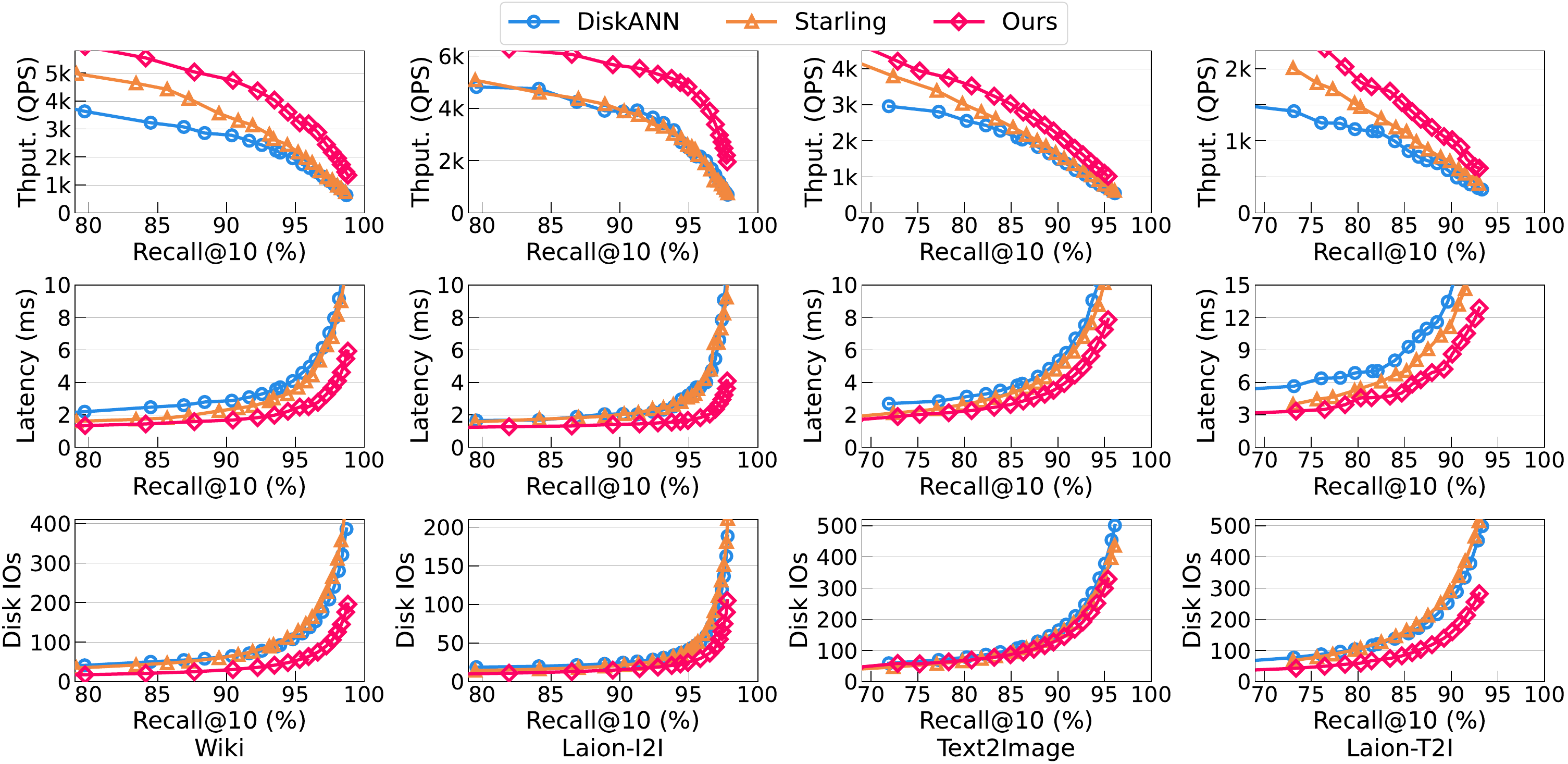}
    \caption{The query throughput, and per query disk IOs of \evname{} and the baselines when using the same memory budget (20\%).}
    \label{fig:eval_main}
\end{figure*}

\subsection{Experiment Settings}
\label{subsec:settings}

\stitlestart{Datasets}
We conducted experiments on the 4 high-dimensional vector datasets in Table~\ref{tab:datasets}. Following PipeRAG~\cite{piperag}, we use the BGE embedding model~\cite{bge} to create a 384-dimensional dataset called \textit{Wiki}, with the 100M base vectors generated from the Wikipedia text data~\cite{wiki} and the 100K query vectors generated from the MMLU text dataset~\cite{mmlu}. The \textit{Text2Image} dataset~\cite{text2image} is a cross-modal dataset with the base vectors derived from images and query vectors derived from texts. Laion~\cite{laion, laion5b} includes two datasets, with dimensions of 512 and 768, respectively, and we use the first 100M image vectors in Laion for both base datasets. For the query vectors, we sample 10K text vectors for the 512-dimensional dataset, and sample 10K image vectors for the 768-dimensional dataset, following RoarGraph~\cite{roargraph} and ANN benchmarks~\cite{annbenchmark}. We donate these two datasets as Laion-T2I and Laion-I2I, respectively. \modify{Existing studies (e.g., ~\cite{bang}) report similar performance trends for 100M and billion-scale datasets. The high-dimensional datasets in our experiments, like Laion-I2I, even consume more space than some billion-scale datasets (SIFT-1B).}



\stitle{Baseline systems}
We compare \evname{} with 2 state-of-the-art systems for disk-based ANNS, i.e., \textit{DiskANN}~\cite{diskann} and \textit{Starling}~\cite{starling}. In particular, \textit{DiskANN} stores the compressed vectors and caches both adjacency lists and vectors in memory. \textit{Starling} builds upon \textit{DiskANN}, conducts graph reordering for improved disk block locality, and stores a small in-memory index for fast navigation. We also create a baseline called \textit{Ours-GR} by replacing the graph-replicated disk layout of \evname{} with Starling's disk layout to evaluate the trade-off of disk space and efficiency. For all systems, following DiskANN, we adopt PQ~\cite{pq} as the vector compression algorithm, and Vamana~\cite{diskann} as the proximity graph index.

We do not compare with SPANN~\cite{spann} as it was outperformed by Starling and DiskANN in~\cite{starling}. Systems that target different settings are also excluded, e.g., AiSAQ~\cite{aisaq} and LM-DiskANN~\cite{lmdiskann} store all data on disk and do not use memory at all; SPFresh~\cite{spfresh} and FreshDiskANN~\cite{freshdiskann} focus on the streaming updates for the vector index; and FusionANNS~\cite{fusionanns} uses GPU to accelerate computation.

\stitle{Testbed}
We conducted all experiments on a server that hosts an Intel(R) Xeon(R) CPU E5-2686 v4 @ 2.30GHz CPU that contains 72 cores and 504GB DRAM. The server is equipped with eight 1.9TB NVMe SSDs, and we applied RAID-0~\cite{raid0} on all its SSDs and obtained a 14TB disk space with an exaggerated bandwidth of 4.0 GB/s. The operating system is Ubuntu 22.04.4 with Linux kernel version 6.8.1. By default, we use a memory that is 20\% of the original dataset, following Starling~\cite{starling}. All systems use the same compression ratio for the compressed vectors, and the experiments were executed with 8 threads, with each query processed by a single thread. 


\stitle{Performance metrics}
Our primary metrics are \textit{query} \textit{throughput}, measured by the average number of queries processed per second (QPS), \textit{query latency}, measured by the average query processing time, and the \textit{average disk IOs}, measured by the average number of disk accesses per query. Following previous works~\cite{diskann, starling, spann}, we use \textit{recall@10} to measure the quality of the search results and report the average recall@10 over the queries.



\begin{table}[!t]
\centering
\setlength{\tabcolsep}{3pt}
\caption{The query throughput and latency of \evname{} and  baseline systems when achieving the target recalls in Table~\ref{tab:datasets}. \textit{Comparison} is the \textit{throughput improvement} and \textit{latency reduction} of \evname{} compared with the best-performing baseline.}
\label{tab:eval_key}
\fontsize{8}{10}\selectfont
\begin{tabular}{c|cccc|cccc}
\toprule[1pt]
& \multicolumn{4}{c|}{\textbf{Throughput (QPS)}} & \multicolumn{4}{c}{\textbf{Latency (ms)}} \\
\midrule
\textbf{Datasets} & \textbf{Wiki} & \textbf{L-I2I} & \textbf{T2Img} & \textbf{L-T2I} & \textbf{Wiki} & \textbf{L-I2I} & \textbf{T2Img} & \textbf{L-T2I} \\
\midrule
\textit{DiskANN} & 1,820 & 2,337 & 1,492 & 589 & 4.39 & 3.42 & 5.36 & 13.56 \\
\textit{Starling} & 2,134 & 2,529 & 1,514 & 691 & 3.74 & 3.16 & 5.28 & 11.56 \\
\evname{} & 3,490 & 4,825 & 2,088 & 1,016 & 2.29 & 1.65 & 3.83 & 8.62 \\
\midrule
Comparison & 1.64x & 1.91x & 1.38x & 1.47x & 61\% & 52\% & 73\% & 75\% \\ 
\bottomrule[1pt]
\end{tabular}
\end{table}

\subsection{Main Results}

Figure~\ref{fig:eval_main} reports the query throughput, query latency, and average disk IOs of \evname{} and the baseline systems. We observe that \evname{} consistently outperforms the baseline systems across all datasets and recall targets. When achieving the same recall, \evname{} improves the query throughput of Starling by 60\% when averaged over datasets, reduces the query latency by 35\%, and reduces the average disk IOs by 39\%. The detailed results are provided in Table~\ref{tab:eval_key}.
Comparing the two baselines, we observe that \textit{Starling} only outperforms \textit{DiskANN} by a small margin. On the Laion-T2I dataset, \textit{Starling} performs almost the same as \textit{DiskANN}. This is because the vector dimension is high for Laion-T2I (i.e., 768), and a 4KB disk block can only store 1 vector, which makes \textit{Starling}'s designs fail. In contrast, \evname{} introduces the graph-replicated disk layout, where a disk block can hold several neighbors' adjacency lists, leading to better disk block utilization and fewer disk accesses.

We also observe that the improvements of \evname{} over the baselines are more pronounced for the single-modal datasets (i.e., Wiki and Laion-I2I) than the cross-modal datasets. This is because the cross-modal datasets require more accurate but larger compressed vectors than the single-modal datasets. 
As such, given the same memory budget, the cross-modal datasets leave less space for the graph cache of \evname{}, which limits its performance gains. Additionally, \evname{} achieves greater improvements on datasets with higher dimensionality. For example, a 91\% throughput improvement for Laion-I2I (the highest-dimensional dataset), compared to a 38\% improvement for Text2Image (the lowest-dimensional dataset).
This is because Text2Image is a cross-modal dataset with a low dimension, less space is reserved for caching graphs compared to other datasets. With a 20\% memory budget, only 25\% of nodes can be stored in the graph cache for Text2Image, whereas more than 60\% of nodes can be cached for the other datasets. Specifically, Laion-I2I can cache the whole index graph in its memory under a 20\% memory budget.

\begin{figure}[!t]
	\centering
	\includegraphics[width=1\linewidth]{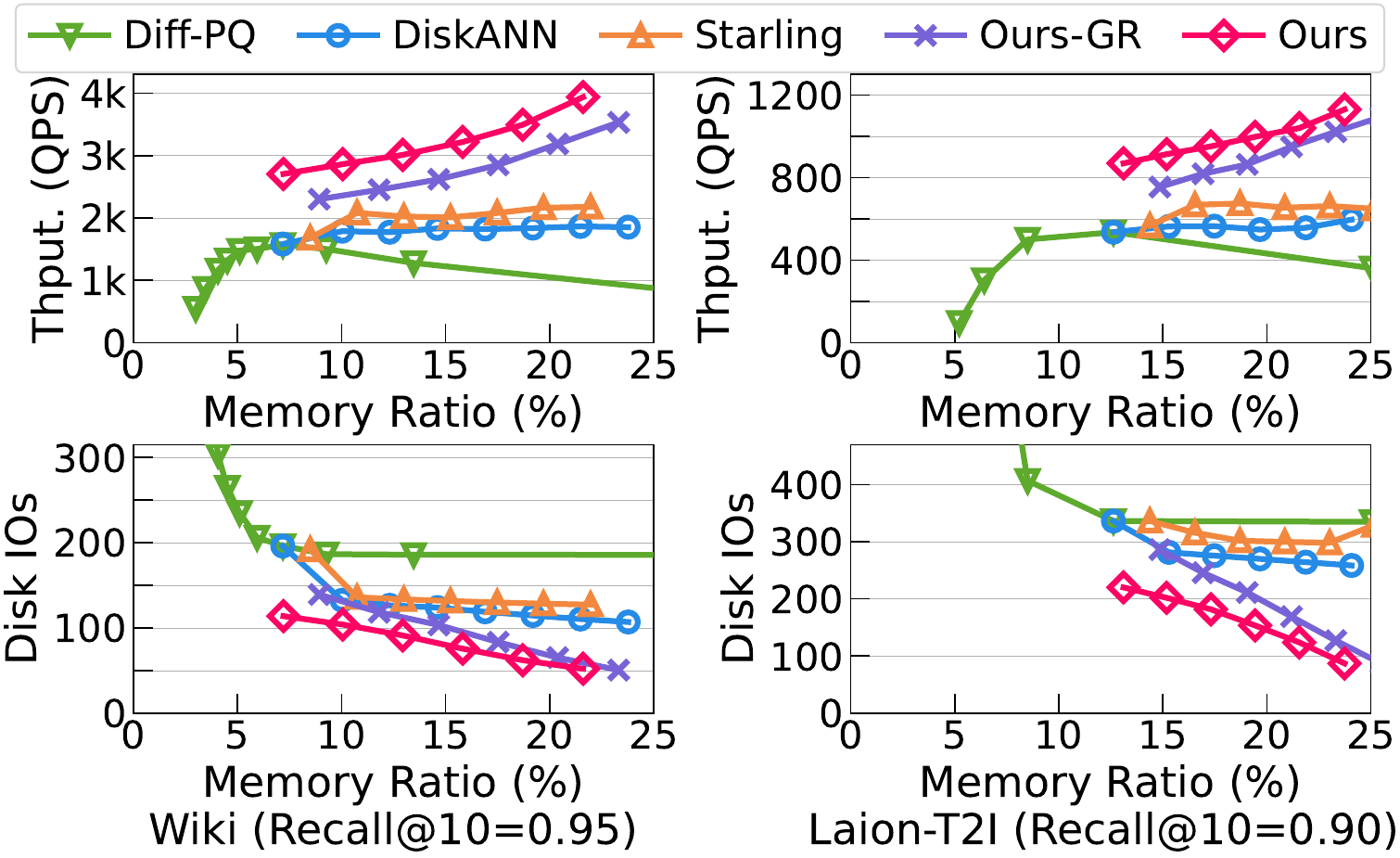}
	\caption{The throughput and Disk IOs of \evname{} and the baselines when varying the memory consumption.}
	\label{fig:eval_membudget}
\end{figure}

\stitle{Varying memory consumption}
Figure~\ref{fig:eval_membudget} reports the query throughput and disk IOs when achieving the recall targets in Table~\ref{tab:datasets} but changing the memory they can utilize. We implement \textit{Diff-PQ} over DiskANN, which adjusts the compression ratios of PQ to use up all memory and does not utilize the memory cache. 
For \textit{Diff-PQ}, we observe that search performance first improves but then decreases with the increase in memory size. This echoes our observation in \S~\ref{subsec:analysis pq}.   
For the other baselines, we adopt the compression ratio that achieves the highest throughput for each dataset.
For \textit{DiskANN} and \textit{Starling}, we observe that large memory only slightly reduces their disk access, and the throughput improvement is negligible. This indicates that both the navigation index and node cache are ineffective under a high memory budget.
For \textit{Ours-GR}, we find that it outperforms the baselines even with the smallest memory budget. This is because it adopts the asynchronous block prefetch and hides the disk access latency. For higher memory budgets, \textit{Ours-GR} has higher throughput because more nodes can be found in the graph cache in the search stage, and disk IOs are saved by checking only the top-ranking nodes in the refinement stage.
Finally, by cooperating with the graph-replicated disk layout, \evname{} achieves higher throughput compared with \textit{Ours-GR}, particularly under lower memory budgets. This is because the layout caches the neighbors' adjacency lists on the disk block, thereby reducing disk access in the search stage.

\begin{figure}[!t]
    \centering
    \begin{minipage}[b]{0.23\textwidth}
        \centering
        \includegraphics[width=\textwidth]{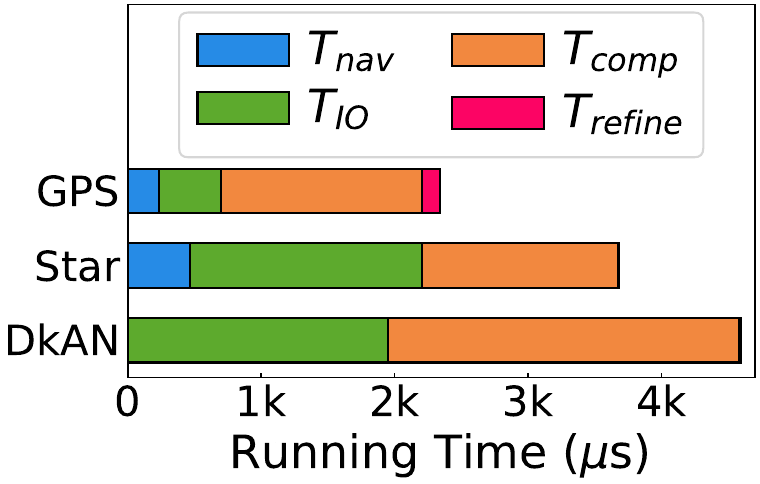}
        \caption{Query time decomposition on the Wiki dataset under the default memory budget and recall target.}
        \label{fig:time_decompose}
    \end{minipage}
    \hfill
    \begin{minipage}[b]{0.23\textwidth}
        \centering
    \includegraphics[width=\textwidth]{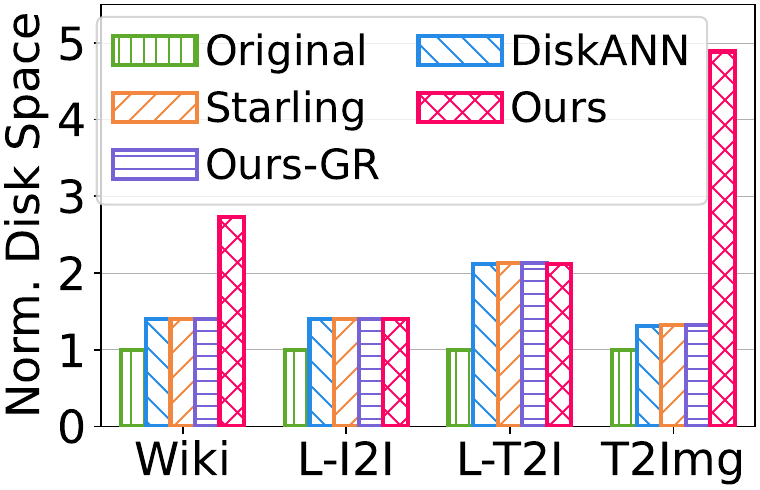}
    \caption{The disk space used by the systems, normalized by the size of the raw \textit{original} vector dataset.}
    \label{fig:eval_disk_space}
    \end{minipage}
\end{figure}

\stitle{Time decomposition}
To understand the speedup of \evname{}, we decompose the request latency of it and the baselines in Figure~\ref{fig:time_decompose}. The dataset is Wiki, and all results were obtained under 95\% recall@10 and 20\% memory budget. $T_{nav}$, $T_{IO}$, $T_{comp}$, and $T_{refine}$ are the time for in-memory index navigation, CPU idling waiting for disk IOs, running the search stage, and refining the final top-$k$, respectively. \evname{}'s $T_{IO}$ includes the waiting time in both the search and refinement stages.
We observe that both \textit{Starling} and \textit{DiskANN} suffer from long disk access time. \textit{Starling} has relatively shorter computation and disk IO time due to the navigation index and execution optimizations. With nearly the same computation time, \evname{} significantly reduces the time for disk IOs using the two-stage search algorithm and asynchronous block prefetch, which reduce the amount of disk IOs and hide the disk IO latency, respectively. Finally, we observe that the execution time for the refinement stage is much shorter than the search stage. This is because the refinement stage computes exact distances only for the top-ranking nodes in $\mathcal{L}{appr}$, which is significantly fewer than the total number of compressed vector computations required for all nodes in $\mathcal{L}{appr}$.

\stitle{Disk space overhead}
Figure~\ref{fig:eval_disk_space} reports the disk space consumption of the systems. \textit{DiskANN}, \textit{Starling}, and \textit{Ours-GR} use the same disk space because they have no additional disk consumption. In contrast, \evname{} introduces a graph-replicated disk layout, which trades disk space for efficiency. For high-dimensional datasets, such as Laion-T2I and Laion-I2I, one 4KB block can only contain one node such that the disk space for \evname{} is the same as the other systems. However, for datasets with relatively lower dimensions, the replicated layout of \evname{} brings larger disk space amplification.

\begin{figure}[!t]
    \centering
    \begin{minipage}[b]{0.23\textwidth}
        \centering
        \includegraphics[width=\textwidth]{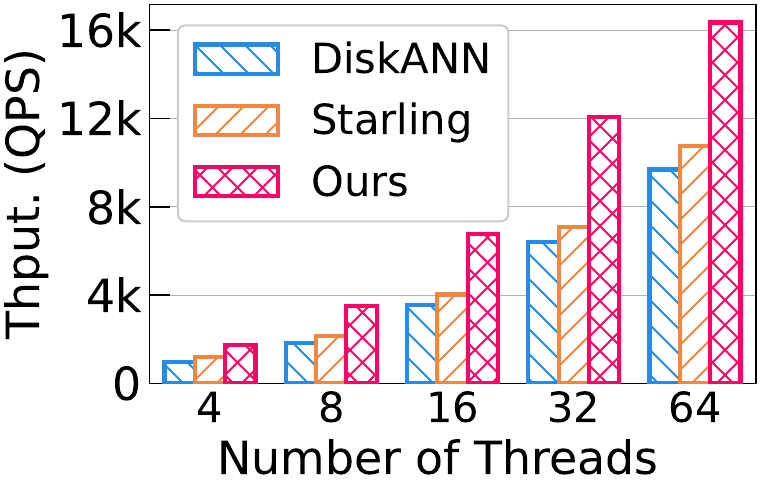}
        \caption{The query throughput of \evname{} and baseline systems when changing the number of threads for query processing.}
        \label{fig:eval_change_thread}
    \end{minipage}
    \hfill
    \begin{minipage}[b]{0.23\textwidth}
        \centering
    \includegraphics[width=\textwidth]{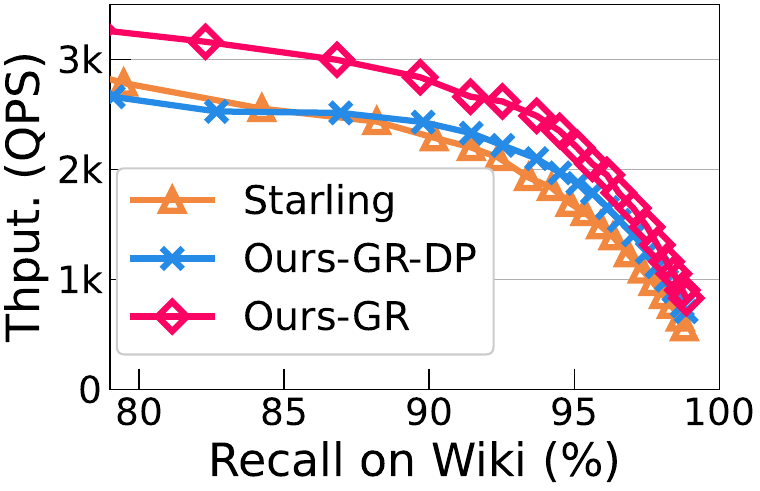}
    \caption{\modify{The effect of \evname{}' asynchronous block prefetch. The memory-resident navigation index and data caches are disabled (i.e., \textit{Ours-GR}).}}
    \label{fig:eval_pipeline}
    \end{minipage}
\end{figure}

\stitle{Varying the number of threads}
Figure~\ref{fig:eval_change_thread} displays the throughput of \evname{} and the baselines when launching varying numbers of threads to process ANNS queries. Other parameters remain the same as the main results. 
We observe that \evname{} consistently outperforms the baselines by a large margin when using different numbers of threads. Specifically, \evname{} achieves an average throughput improvement of 83\% over \textit{DiskANN} and 60\% over \textit{Starling}, respectively. The reason is that \evname{} requires less disk access for each query, rendering its performance less dependent on the disk bandwidth when using many threads.

\subsection{Micro Experiments}
\label{subsec:designchoices}

\begin{figure}[!t]
    \centering
\includegraphics[width=1\linewidth]{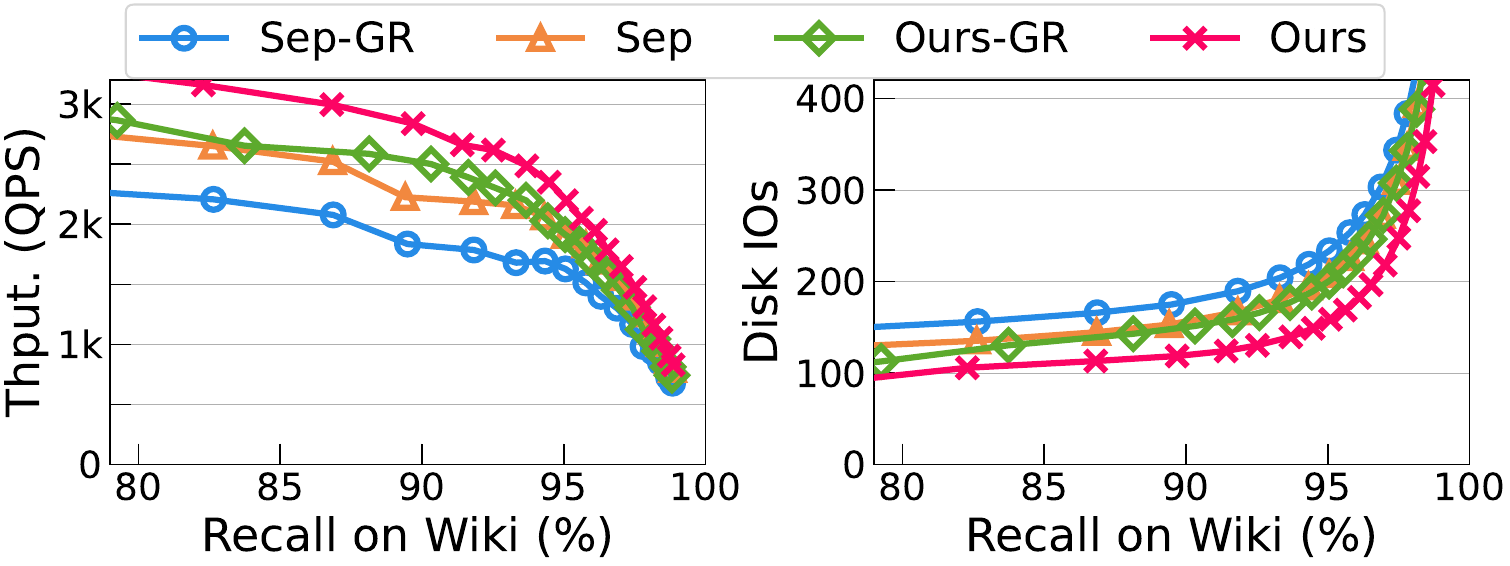}
\caption{Comparing the \evname{}' graph replicated disk layout with the vector-graph separation layout.}
\label{fig:eval_sep}
\end{figure}




\stitle{Asynchronous block prefetch}
To evaluate the gain of our asynchronous block prefetch, we create a baseline \textit{Ours-GR-DP}, which disables the block prefetch. Figure~\ref{fig:eval_pipeline} compares the query throughput. Compared with \textit{Ours-GR-DP}, \textit{Ours-GR} achieves a throughput improvement of 18\% for the Wiki dataset under 95\% recall. This indicates that the block prefetch does not affect query accuracy, and it effectively hides the latency of disk access, thus improving the system performance.



\stitle{Alternatives for the disk layout}
For each disk block, our graph-replicated disk layout places a node's adjacency list, exact vector, and its neighbors' adjacency lists to increase the disk block data locality. Two alternative layouts that may also improve disk data locality are described in \S~\ref{subsec:disk block}. The first is the vector-graph separation layout~\cite{separation}, which classified the blocks into graph blocks and vector blocks, with graph blocks containing only adjacency lists and vector blocks containing only vectors(i.e., Figure~\ref{fig:layout}(b)).
We created two baselines based on it, i.e., \textit{Sep-GR} and \textit{Sep}. Particularly, \textit{Sep-GR} applies \textit{Starling}'s re-layout strategy to both the graph blocks and vector blocks; while \textit{Sep} trades disk space for efficiency, replicating a node's neighbors' adjacency lists in the graph blocks. Compared with \textit{DiskANN}'s disk consumption, on the Wiki dataset, \textit{Sep-GR} requires an additional 10\% disk space for block padding, while \textit{Sep} consumes an extra 200\% disk space for graph replication.

Figure~\ref{fig:eval_sep} compares the throughput and disk IO between the separation layout and \evname{}'s layout, using the same two-stage search algorithm. The results suggest that the graph-vector separation layout requires more disk IOs and results in lower throughput compared to \evname{}'s layout. The underlying reason is that, in the refinement stage of the separation layout, all top-ranking vectors need to be loaded from the disk and each of them requires one disk access, as none of them are accessed during the search stage. This introduces an extra disk IO, and two disk accesses are needed for some nodes. \modify{As a result, the separation layout leads to poorer performance due to the additional I/O operations. Even when employing the graph-replicated layout where each graph block can store the adjacency lists of 20 neighbors (\textit{Sep}), the performance remains inferior to that of \evname{}.
 }

\begin{figure}[!t]
    \centering
\includegraphics[width=1\linewidth]{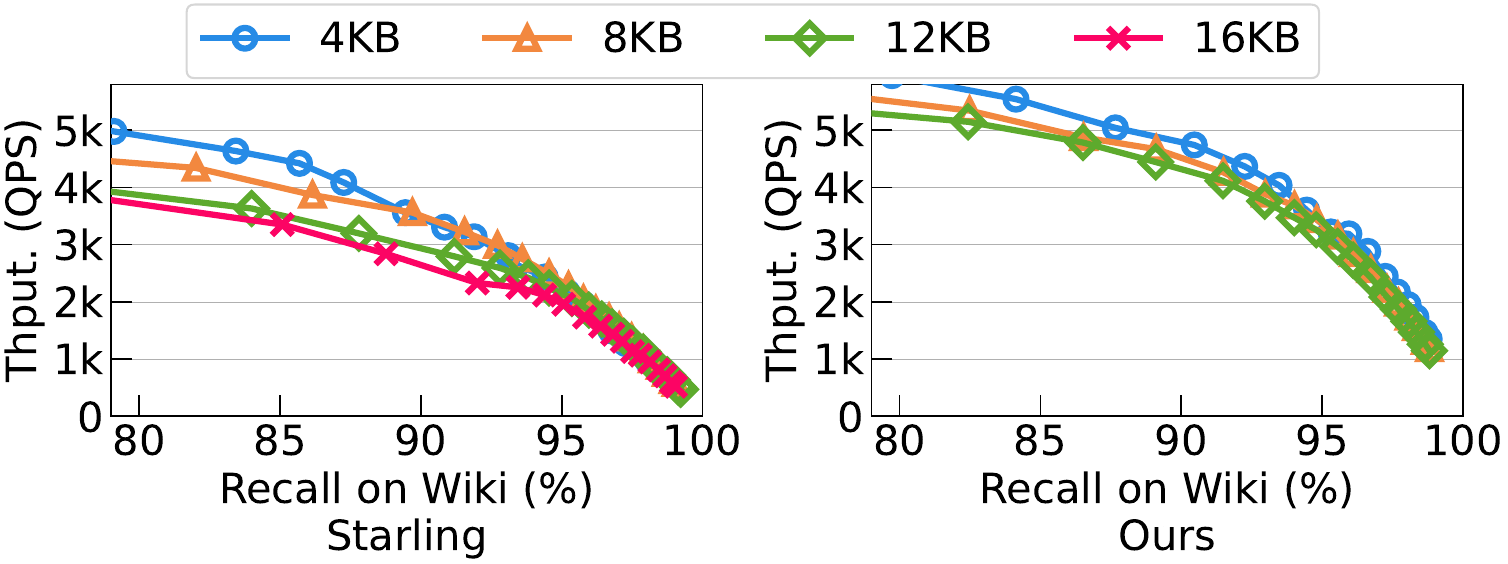}
\caption{The query throughput of Starling and \evname{} when using 20\% memory budget but varying the block size.}
\label{fig:eval_change_page_size}
\end{figure}

The second alternative is using larger block sizes (i.e., Figure~\ref{fig:layout}(c)).
Figure~\ref{fig:eval_change_page_size} reports the query throughput of \textit{Starling} and \evname{} when changing the block sizes. With a larger block size, a single block can accommodate more nodes in \textit{Starling} or replicate more neighbors in \evname{}, while the cost of reading a block also increases. Note that a 12KB block size is sufficient to store all neighbors within a graph-replicated block in \evname{}. We observe that system throughput slightly decreases as the block size increases for both \textit{Starling} and \evname{}. This is because a larger block size consumes more disk bandwidth for each disk access.



\begin{figure}[!t]
    \centering
\includegraphics[width=1\linewidth]{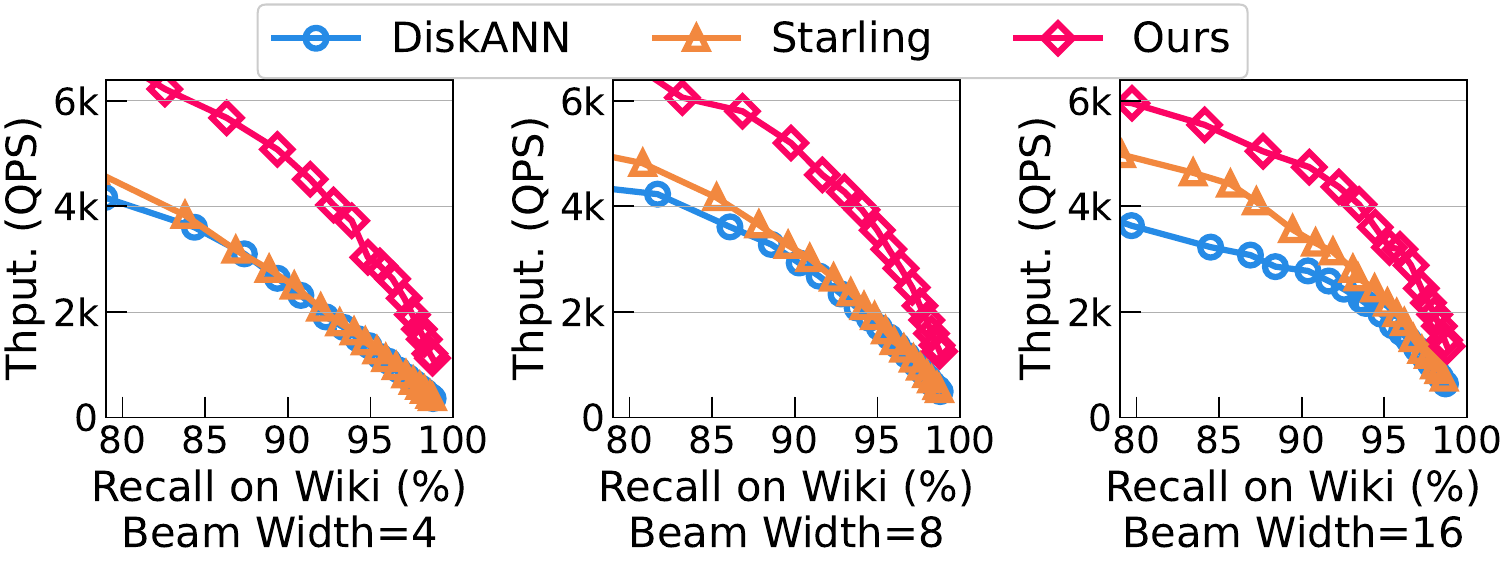}
\caption{The query throughput of \evname{} and baselines when using 20\% memory budget but changing the beam width.}
\label{fig:eval_change_bw}
    \vspace{-3mm}
\end{figure}

\stitle{Changing beam widths}
Figure~\ref{fig:eval_change_bw} reports the throughput of \evname{} and the baselines when using different beam widths. In particular, beam width controls the number of candidate nodes checked in each iteration, and a larger beam width allows a larger disk read batch but may also introduce extra disk access, especially when the queue size  $\mathcal{D}$ is small. The results show that \evname{} outperforms \textit{DiskANN} and \textit{Starling} across all beam widths. Notably, \evname{} exhibits consistent performance across different beam widths, while \textit{Starling} achieves better throughput with larger beam widths. This is because \evname{}' asynchronous block prefetch hides a part of the disk access latency, while other baselines may suffer from longer disk access latency under smaller beam widths.

\section{Related Work}

\stitle{ANNS indexes} 
There are two main types of indexes for ANNS, i.e., IVF~\cite{ivf} and proximity graph~\cite{hnsw}. IVF clusters the vectors and only checks a few nearest clusters for each query. \modify{Many IVF-based index structures are proposed with different clustering methods and search algorithms~\cite{ivf, bang, fusionanns, spann, juno, micronn}.} Proximity graph connects similar vectors to form a graph and processes queries by graph traversal. Proximity graph provides the state-of-the-art performance for ANNS, achieving the same recall with much fewer distance computations than IVF~\cite{hnsw}. As such, both DiskANN~\cite{diskann} and Starling~\cite{starling} adopt proximity graphs. There are many variants for proximity graph~\cite{hnsw, nsg, knng, roargraph, pqbuild, rdesent, parlayann, randnsg, lshapg, lira, leann}, they differ in the graph construction and edge connection rules but adopt the same graph traversal procedure for query processing. Thus, their data access patterns are the same for disk-based ANNS, and \name{} can be generalized across them. The vector database in ANNS indexes could be embedded from language models~\cite{bge, ali_search}, image models~\cite{ann4text_retrieval, taobao_retrival}, recommendation models~\cite{fbdlrm, carina}, or graph models~\cite{ge2, dgi, atom}.

Vector quantization is an important technique for ANNS that compresses each vector for smaller storage and faster distance computation. There are many vector quantization methods, e.g.,  PQ~\cite{pq}, OPQ~\cite{opq}, LVQ~\cite{lvq}, deltaPQ~\cite{deltapq}, and RabitQ~\cite{rabitq, extendpq}, that work in different ways. DADE~\cite{dade}, GleanVec~\cite{gleanvec}, LeanVec~\cite{leanvec}, and ADSampling~\cite{adsampling} utilize projection methods like PCA to reduce the vector dimension for acceleration. \name{} generalizes across the quantization methods because they are used as a black box to compute approximate distance. \modify{Some researches consider more complex vector queries than top-$k$ nearest neighbors, e.g., ANNS with range or attribute filters~\cite{serf, irangegraph, filterdiskann, unify, digra}. However, proximity graph is also popular for these queries, and thus \name{} can be extended by incorporating the attributes.}

\stitlestart{Disk-based ANNS} To handle large vector datasets at a low cost, several systems use disk for data storage. Both DiskANN~\cite{diskann} and Starling~\cite{starling} run with a small memory (e.g., 10-20\% of the dataset size) and large disk. \name{} targets the same scenario but outperforms them by a large margin by carefully designing the data layout to reduce disk IOs. SPANN~\cite{spann} and FusionANNS~\cite{fusionanns} adopt the IVF index for disk-based ANNS. SPANN stores the cluster centers in memory to identify the nearest centers for a query and fetch the required clusters (of vectors) from the disk on demand. FusionANNS keeps the cluster centers on CPU memory, compressed vectors on GPU memory, and exact vectors on disk. A query first identifies its nearest clusters on the CPU, then uses GPU to compute approximate distances for (the vectors in) these clusters, and finally reads the top-ranking vectors from disk for re-ranking with exact distance. Since the IVF index computes more distances than proximity graph index, SPANN is outperformed by Starling~\cite{starling}, and small GPU memory limits the accuracy of the compressed vectors for FusionANNS, which may require reading many exact vectors for re-ranking. LM-DiskANN~\cite{lmdiskann} and AiSAQ~\cite{aisaq} target a more extreme scenario that uses only disk but no memory. 

\stitle{Other ANNS systems}
Some studies handle large vector datasets using emerging hardware, such as HM-ANN~\cite{hmann} for persistent memory, CXL-ANNS~\cite{cxlanns} for CXL memory, MemANNS~\cite{memanns} for UPMEM PIM, and SmartANNS~\cite{smartssdann} for SmartSSD. These systems are tailored to fit the performance characteristics of their target hardware. To improve search efficiency, some works employ accelerators such as GPUs~\cite{fusionanns, bang, ggnn, ganns, cagra, pilotann} and FPGAs~\cite{fpgaanns, dfgas, fanns} to accelerate the distance computations during ANNS. Several studies scale ANNS beyond a single server. For example, Milvus~\cite{milvus}, AnalyticDB~\cite{adb}, Hakes~\cite{hakes}, and d-HNSW~\cite{dhnsw} can run ANNS over multiple distributed machines, while Vexless~\cite{vexless} utilizes serverless cloud functions.

\section{Conclusion}

In this paper, we revisit the storage layout for disk-based graph indexes.
Through extensive profiling and analysis, we observe that for proximity graph, which is the state-of-the-art index for vector search, the adjacency lists are more important than the vectors.
Guided by this insight, we prioritize the adjacency lists over the vectors and propose two key storage layout designs, i.e., graph prioritized memory cache and graph replicated disk block.
Based on the new storage layout, we built an efficient and general disk-based ANNS \name{} system, and \name{} significantly improves the performance of state-of-the-art systems for disk-based vector search.



\bibliographystyle{ACM-Reference-Format}
\bibliography{sample}


\end{document}